\documentclass[epj,onecolumn]{webofc}
\usepackage[varg]{txfonts} 
\usepackage{wasysym}
\usepackage{pifont}
\usepackage{wrapfig}        
\woctitle{KLOE-2 Workshop on $e^+e^-$ collider physics at 1 GeV}
\newcommand{\ha}{\frac12 }

\newcommand{\dalh}{\Delta \alpha^{\rm had}}
\newcommand{\cin}[1]{ #1}
\newcommand{\gv}{\mbox{GeV}}
\newcommand{\mv}{\mbox{MeV}}
\newcommand{\ppm}{\pi^+ \pi^-}

\newcommand{\bl}{\phantom{-}}

\newcommand{\mbo}[1]{$#1$ }
\newcommand{\epm}{e^+e^- }
\newcommand{\epem}{e^+e^- }
\newcommand{\power}[1]{\times 10^{#1} }
\newcommand{\I}{\rm i }
\newcommand{\D}{\rm d }
\newcommand{\E}{\rm e }
\newcommand{\Impa}{\rm Im }
\newcommand{\Repa}{\rm Re }
\newcommand{\semis}{\;;\;\; }
\newcommand{\amu}{a_\mu }
\newcommand{\amuh}{a_\mu^{\rm had} }
\newcommand{\bea}{\begin{eqnarray}}
\newcommand{\eea}{\end{eqnarray}}
\newcommand{\epo}{\;. }

\newcommand{\bary}{\begin{array}}
\newcommand{\eary}{\end{array}}

\newcommand{\ttc}[1]{\multicolumn{2}{c}{#1}}
\newcommand{\amuexp}{a_\mu^{\mathrm{exp}}}
\newcommand{\amuthe}{a_\mu^{\mathrm{the}}}
\newcommand{\gapprox}{\apprge}
\newcommand{\lapprox}{\apprle}

\begin{document}
\title{Muon $g-2$ theory: the hadronic part}

\author{\firstname{Fred} \lastname{Jegerlehner}\inst{1,2}\fnsep\thanks{\email{fjeger@physik.hu-berlin.de}}
}

\institute{
Deutsches  Elektronen--Synchrotron (DESY), Platanenallee 6, D--15738 Zeuthen, Germany
\and
Humboldt--Universit\"at zu Berlin, Institut f\"ur Physik, Newtonstrasse 15, D--12489 Berlin,
Germany
          }

\abstract{%
I present a status report of the hadronic vacuum polarization effects
for the muon $g-2$, to be considered as an update
of~\cite{Jegerlehner:2015stw}. The update concerns recent new
inclusive $R$ measurements from KEDR in the energy range 1.84 to 3.72
GeV. For the leading order contributions I find
$\amu^{\mathrm{had}(1)}=(688.07\pm 4.14)[688.77\pm3.38]\times
10^{-10}$ based on $\epm$data [incl. $\tau$ data],
$\amu^{\mathrm{had}(2)}= (-9.93\pm 0.07) \times 10^{-10}$ (NLO) and
$\amu^{\mathrm{had}(3)}= (1.22\pm 0.01) \times 10^{-10}$ (NNLO).
Collecting recent progress in the hadronic light-by-light scattering
I adopt $\pi^0,\eta,\eta'$ [$95 \pm 12$] + axial--vector [$8 \pm
~3$] + scalar [$-6\pm ~1$] + $\pi,K$ loops [$-20\pm 5$] + quark loops
[$22\pm ~4$] + tensor [$1\pm ~0$] + NLO [$3\pm ~2$] which yields
$ a^{(6)}_\mu(\mathrm{lbl},\mathrm{had})=(103 \pm 29) \power{-11}.$
With these  updates I find
$a_\mu^{\rm exp}-a_\mu^{\rm the}=(31.3\pm 7.7)\times 10^{-10}$ a 4.1
$\sigma$ deviation. Recent lattice QCD results and future prospects
to improve hadronic contributions are discussed.}
%
\thispagestyle{empty}
\begin{flushright}
DESY 17-058,~~HU-EP-17/12\\
April 2017
\end{flushright}

\vfill

\begin{center}
{\large\bf
Muon $g-2$ Theory: the Hadronic Part}\\
{Fred Jegerlehner}\\
{Deutsches  Elektronen--Synchrotron (DESY), Platanenallee 6,\\ D--15738 Zeuthen, Germany\\
Humboldt--Universit\"at zu Berlin, Institut f\"ur Physik, Newtonstrasse 15,\\ D--12489 Berlin,
Germany}

\vfill

\begin{minipage}{0.8\textwidth}
{\bf Abstract}\\
I present a status report of the hadronic vacuum polarization effects
for the muon $g-2$, to be considered as an update
of~\cite{Jegerlehner:2015stw}. The update concerns recent new
inclusive $R$ measurements from KEDR in the energy range 1.84 to 3.72
GeV. For the leading order contributions I find
$\amu^{\mathrm{had}(1)}=(688.07\pm 4.14)[688.77\pm3.38]\times
10^{-10}$ based on $\epm$data [incl. $\tau$ data],
$\amu^{\mathrm{had}(2)}= (-9.93\pm 0.07) \times 10^{-10}$ (NLO) and
$\amu^{\mathrm{had}(3)}= (1.22\pm 0.01) \times 10^{-10}$ (NNLO).
Collecting recent progress in the hadronic light-by-light scattering
I adopt $\pi^0,\eta,\eta'$ [$95 \pm 12$] + axial--vector [$8 \pm
~3$] + scalar [$-6\pm ~1$] + $\pi,K$ loops [$-20\pm 5$] + quark loops
[$22\pm ~4$] + tensor [$1\pm ~0$] + NLO [$3\pm ~2$] which yields
$ a^{(6)}_\mu(\mathrm{lbl},\mathrm{had})=(103 \pm 29) \power{-11}.$
With these  updates I find
$a_\mu^{\rm exp}-a_\mu^{\rm the}=(31.3\pm 7.7)\times 10^{-10}$ a 4.1
$\sigma$ deviation. Recent lattice QCD results and future prospects
to improve hadronic contributions are discussed.
\end{minipage}
\end{center}
\vfill
\noindent\rule{8cm}{0.5pt}\\
$^*$ Invited talk
 KLOE-2 Workshop on $e^+e^-$ collider physics at 1 GeV''
26-28 October 2016 INFN - Laboratori Nazionali di Frascati, Italy
\setcounter{page}{0}
\newpage

\maketitle
\section{Overview: hadronic effects in $g-2$.}
\label{intro}
This review of the hadronic vacuum polarization (HVP) contributions to the muon $g-2$ is to be
considered as a complement to the theory reviews by Marc Knecht and Massimiliano Procura
which focus on the hadronic light-by-light (HLbL) part and the
reviews on hadronic cross sections by Graziano Venanzoni, Simon
Eidelman and Achim Denig in these Proceedings.

\begin{figure}[h]
\centering
\includegraphics[width=0.96\textwidth]{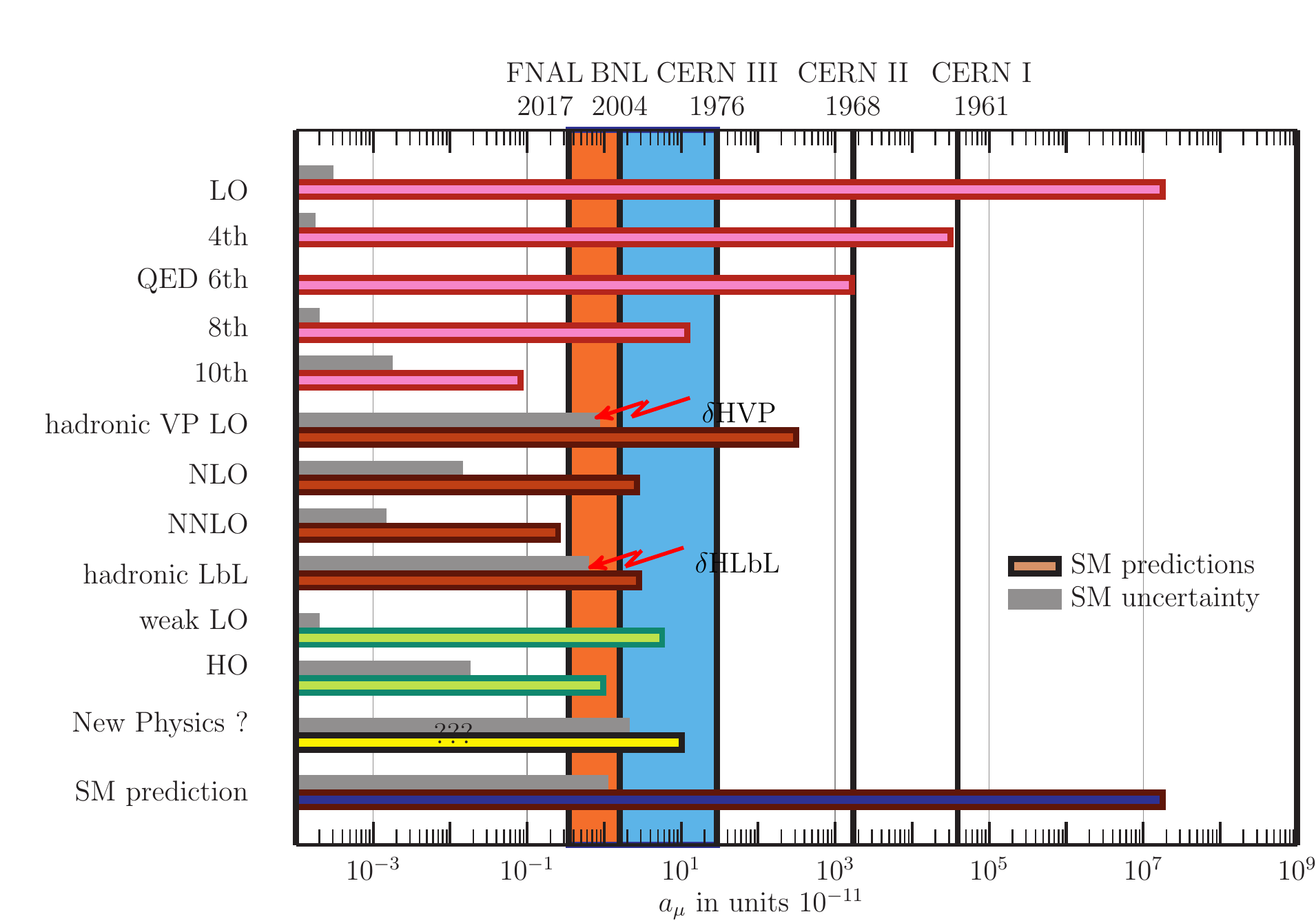}
\caption{Past and future $g-2$ experiments testing various
contributions. As New Physics $?$ we display the
deviation $
(a_\mu^\mathrm{exp}-a_\mu^\mathrm{the})/a_\mu^\mathrm{exp}$. Arrows
point to what is limiting theory precision presently: the Hadronic Vacuum
Polarization (HVP) and Hadronic Light-by-Light (HLbL) contributions.}
\label{fig:contributions17}
\end{figure}

The present experimental muon $g-2$ result from Brookhaven (BNL)
$\amuexp=(11\,659\,209.1\pm5.4\pm3.3[6.3])\power{-10}$~\cite{BNLfinal} soon will be
improved by the new muon $g-2$ experiments at Fermilab and J-PARC. The
Fermilab experiment will be able to reduce the error by a factor 4, the
J-PARC experiment will provide an important cross check with a very
different technique~\cite{Hertzog15}. It means that
the new muon $g-2$ experiments are expected to establish a possible new
physics contribution at the level $\Delta \amu=\amuexp-\amuthe=6.7\,
\sigma$ provided theory remains as it is today and the central value does
not move significantly. If we achieve a reduction of the hadronic
uncertainty by factor 2 we would arrive at $
\Delta \amu =11.6\,\sigma$. That's what we hope to
achieve. Figure~\ref{fig:contributions17} illustrates the present
status and what has been achieved so far.
The present results $a_\mu^{\rm HVP \ LO}  = (6888\pm34) \times 10^{-11}$ amounts to +59.09
$\pm$0.30 ppm, which poses the major challenge.  The subleading
results $a_\mu^{\rm HVP \ NLO}= (-99.3\pm 0.7) \times 10^{-11}$ and
$a_\mu^{\rm HVP \ NNLO} = (12.2\pm 0.1) \times 10^{-11}$ although
relevant will be known well enough. These number also compare with the
well established weak $a_\mu^{{\rm EW}}= (154\pm 1) \times 10^{-11}$
at 1.3 $\pm$ 0.0 ppm
and the problematic HLbL estimated to contribute $a_\mu^{{\rm
HLbL}}= (103 \pm 29~[105\pm26])\times 10^{-11}$, which is
representing a +0.90 $\pm$0.25 ppm effect.\\

\begin{figure}
\centering
\includegraphics[height=3cm]{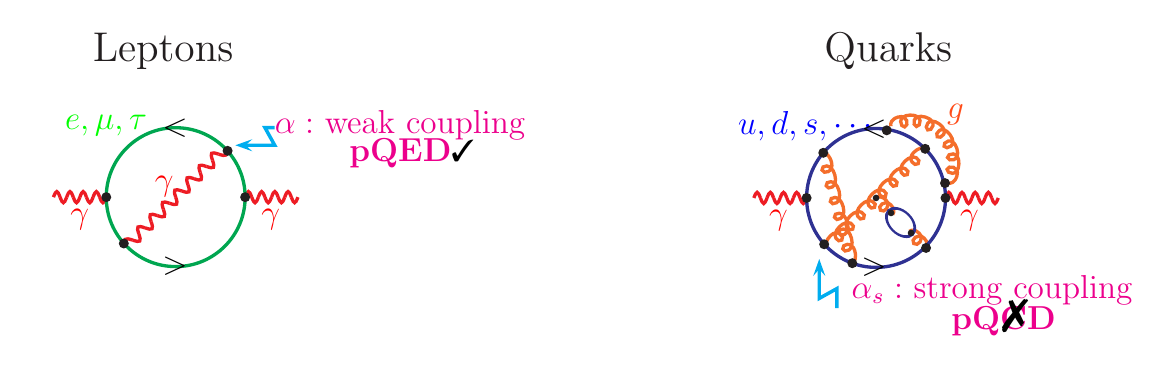}\\[-1cm]
\includegraphics[height=3cm]{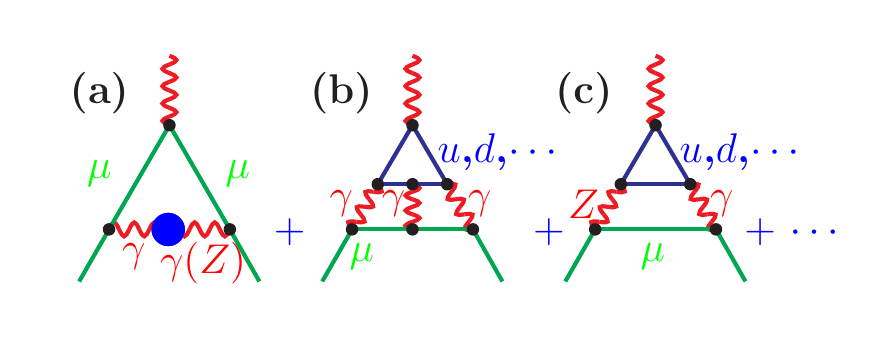}
\caption{In the upper panel we compare leptonic with hadronic vacuum
polarization effects. The lower panel illustrates the three classes of
$g-2$ contributions exhibiting substantial hadronic corrections.}
\label{fig:hadstuff}
\end{figure}

Virtual effect form low energy hadronic excitations are the standard
problem in electroweak precision physics. At a certain level of
precision predictions are hampered by non-perturbative effects, which
technically are not under desirable control on the theory side. For
the muon $g-2$ the leading hadronic effects are related to the
diagrams in figure~\ref{fig:hadstuff}
and concern
\begin{tabular}{l}
(a)~Hadronic vacuum polarization (HVP) of order $O(\alpha^2),O(\alpha^3)$, \\
(b)~Hadronic light-by-light scattering (HLbL) of order $O(\alpha^3)$, \\
(c)~Hadronic effects in 2-loop hadronic electroweak (HEW) corrections
of order sub--$O(\alpha G_F m_\mu^2)$.
\end{tabular}

\noindent
Light quark loops appear as non-perturbative hadronic ``blobs''. The
evaluation of the corresponding non-perturbative effects relies on
hadron production data in conjunction with Dispersion Relations (DR),
or on low energy effective modeling by the Resonance Lagrangian
Approach (RLA), specifically by the Hidden Local Symmetry (HLS) model~\cite{HKS95},
or the Extended Nambu--Jona-Lasinio (ENJL) model~\cite{BPP1995},
large--$N_c$ QCD inspired methods~\cite{KnechtNyffeler01} and on lattice
QCD. Different strategies apply for the different kinds of
contributions:

(a) HVP one evaluates via a dispersion integral over \mbo{\epem \to
\mathrm{hadrons}} data. Here 1 independent amplitude is to be determined by
one specific data set. Global fits based on the RLA (like HLS) allow
to improve the data-driven evaluations~\cite{Benayoun:2011mm}. Lattice
QCD is the ultimate tool to get QCD predictions in future.

(b) HLbL so far has been evaluated by modeling via the Resonance
Lagrangian Approach (RLA) (chiral perturbation theory (CHPT) extended
by vector meson dominance (VMD) in accord with chiral structure of
QCD) or by large--$N_c$ inspired methods and operator product
expansions (OPE). A data driven approach based on dispersion
relations~\cite{Colangelo:2014pva} is attempting to exploit
\mbo{\gamma\gamma \to
\mathrm{hadrons-data}} systematically (here 19 independent amplitudes
are to be determined by as many independent data sets, fortunately not
all are equally important numerically). Also in this case lattice
QCD for me is the ultimate approach, although tough to be achieved
with limited computing resources.

(c) HEW corrections due to quark triangle diagrams: since triple
vector amplitudes vanish $VVV = 0$ by Furry's theorem only $VVA$ (of
\mbo{f\bar{f}Z}-vertex) contributes. Thus it is ruled by the ABJ
anomaly, which is perturbative and non-perturbative simultaneously,
i.e. the leading effects are calculable. The anomaly cancellation condition
intimately relates quark and lepton contributions and the
potentially large leading corrections cancel~\cite{KPPdeR02,MV03,CMV03}
such that hadronic corrections are well under control.

\section{Evaluation of the leading order $a_\mu^{\rm had}$}
\begin{wrapfigure}{l}[3pt]{2.6cm}
\includegraphics[height=2cm]{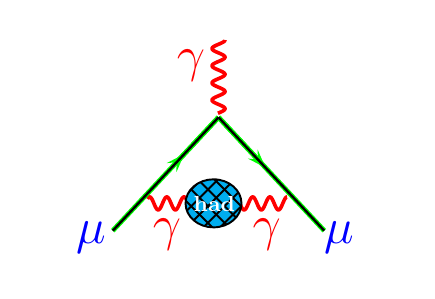}
\end{wrapfigure}
\begin{figure}[h]
\vspace*{-6mm}
\centering
\includegraphics[height=6cm]{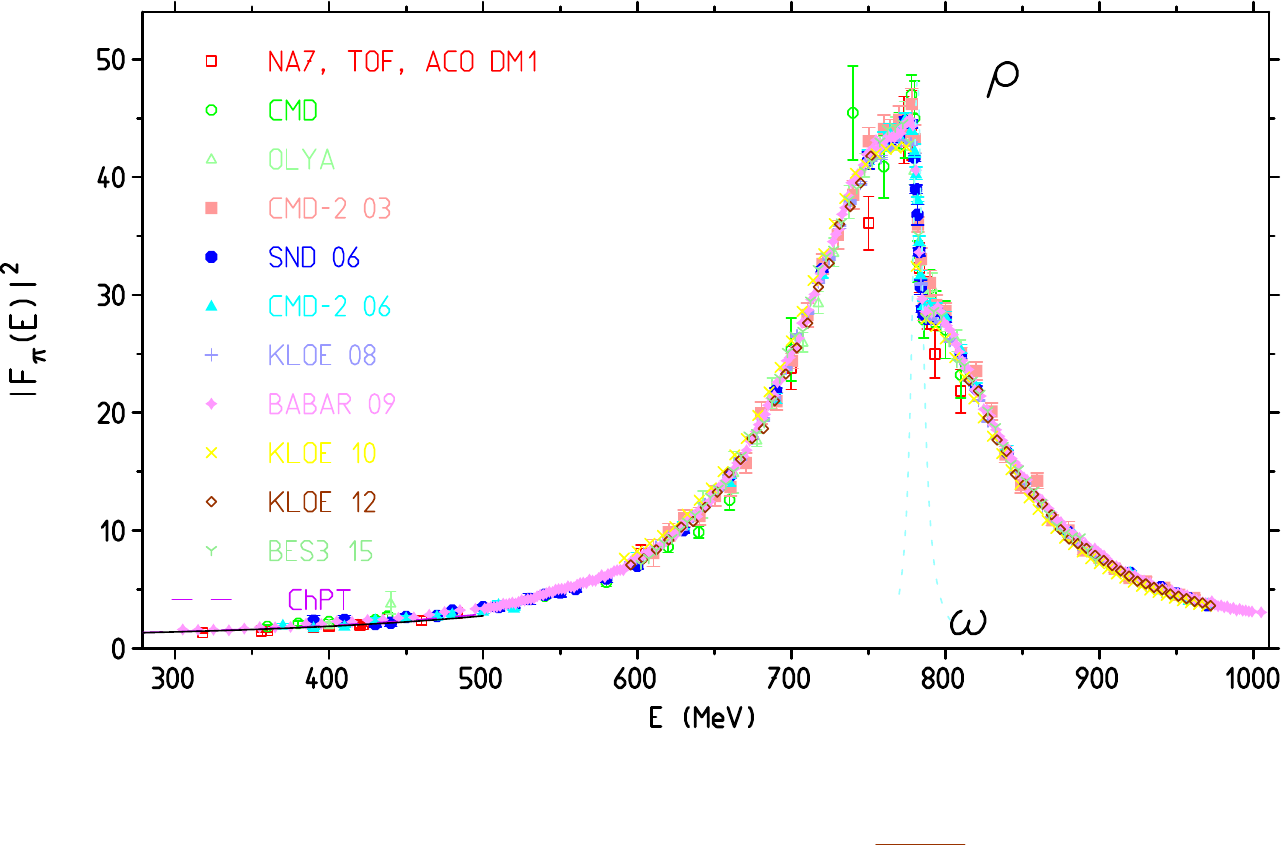}
\caption{A compilation of the modulus square of the pion form factor in the $\rho$ meson region,
which yields ~~~~ about 75\% of $\amuh$. The corresponding $R(s)$ is
$R(s)=\frac14\,\beta_\pi^3\,|F_\pi^{(0)}(s)|^2\,,\,\,\beta_\pi=\sqrt{1=4m^2_\pi/s}$
is the pion velocity.}
\label{fig:VPdiadata}
\end{figure}

The hadronic contribution to the vacuum polarization can be evaluated,
with the help of dispersion relations, from the energy scan of the
ratio $R_{\gamma}(s)\equiv \sigma^{(0)}(e^+e^-\to \gamma^*\to {\rm
hadrons}) /\frac{4\pi\alpha^2}{3s}$ which can be measured up to some
energy $E_{\rm cut}$ above which we can safely use perturbative QCD
(pQCD) thanks to asymptotic freedom of QCD. We apply pQCD from 5.2~GeV
to 9.46~GeV and above 11.5~GeV (see figure~\ref{fig:Rofs} below). Note that the DR requires the
undressed (bare) cross--section $\sigma^{(0)}(e^+e^-\to \gamma^*\to {\rm
hadrons})=\sigma(e^+e^-\to \gamma^*\to {\rm
hadrons})\,|\alpha(0)/\alpha(s)|^2$. The lowest order (LO) VP
contribution is given by
\bea
\amuh = \left(\frac{\alpha m_\mu}{3\pi}
\right)^2 \bigg(\;\;\;
\int\limits_{m_{\pi^0}^2}^{E^2_{\rm cut}}ds\,
\frac{{ R^{\mathrm{data}}_\gamma(s)}\;\hat{K}(s)}{s^2}
+ \int\limits_{E^2_{\rm cut}}^{\infty}ds\,
\frac{{R^{\mathrm{pQCD}}_\gamma(s)}\;\hat{K}(s)}{s^2}\,\,
\bigg)\,,~~~~~~~~~~~~~
\label{amuDRbasic}
\eea
where $\hat{K}(s)$ is a known kernel function growing form $0.63\cdots$ at the
$2m_\pi$ threshold to 1 as $s\to \infty$. The integral is dominated by
the $\rho$ resonance peak shown in figure~\ref{fig:VPdiadata}.
\begin{figure}
\centering
\includegraphics[height=4.5cm]{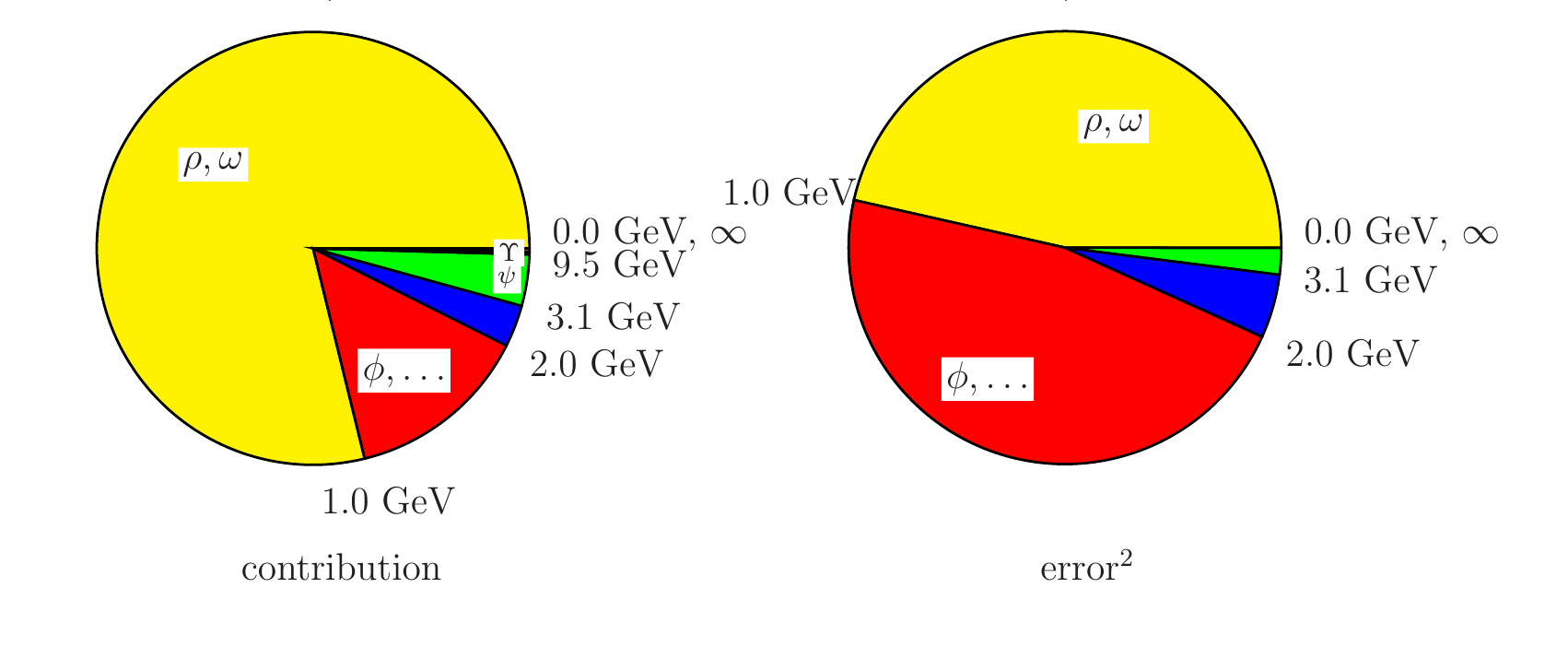}
\caption{Distribution of contributions and error squares from different
energy regions.}
\label{fig:gm2dist}
\end{figure}
The experimental errors imply the dominating theoretical
uncertainties. As a result I obtain
\begin{equation}
\amu^{\mathrm{had}(1)}=(688.07\pm 4.14)[688.77\pm3.38]\:10^{-10}\semis
\epm-{\rm data \ based \ [incl.\ } \tau ]\epo
\label{LOHVP}
\end{equation}
Figure~\ref{fig:gm2dist} shows the distribution of contributions and
errors between different energy ranges.
One of the main issues is
$R_\gamma(s)$ in the region from 1.2~GeV to 2.0~GeV, where more than
30 exclusive channels must be measured and although it contributes
about 20\% only of the total it contributes about 50\% of the uncertainty.
In the low energy region, which is particularly important for the dispersive
evaluation of the hadronic contribution to the muon $g-2$, data have improved
dramatically in the past decade for the dominant $e^+e^- \to
\pi^+\pi^-$ channel (CMD-2~\cite{CMD203}, SND/Novosibirsk~\cite{SND06},
KLOE/Frascati~\cite{KLOE08,KLOE10,KLOE12},
BaBar/SLAC~\cite{BABARpipi},
BES-III/Beijing~\cite{BESIII}) and the statistical errors are a minor
problem now. Similarly, the important region between 1.2 GeV to 2.4 GeV
has been improved a lot by the BaBar exclusive channel measurements in the ISR
mode~\cite{BaBar05,BaBar11,Davier:2015bka,Davier:2016udg}.
Recent data sets collected are: $e^+e^-\to 3(\pi^+\pi^-)$, $e^+e^-\to
\bar{p}p$ and $e^+ e^- \to K^0_{S}K^0_{L},K^+K^-$ from
CMD-3~\cite{Akhmetshin:2013xc,Kozyrev:2016raz},
and $e^+e^-\to \bar{n}n$, $e^+e^-\to \eta \pi^+\pi^-$, $e^+e^-\to
\pi^0\gamma$, $e^+e^- \to \omega\eta\pi^0$, $e^+e^- \to \omega\eta$,
$e^+e^- \to K^+K^-$ and $e^+e^- \to \omega\pi^0 \to \pi^0\pi^0\gamma$ from
SND~\cite{Achasov:2014ncd,Aulchenko:2014vkn,Achasov:2016bfr}.

Above 2~GeV fairly accurate BES-II data~\cite{BES02} are
available. Recently, a new inclusive determination of $R_\gamma(s)$ in
the range 1.84 to 3.72 GeV has been obtained with the KEDR detector at
Novosibirsk~\cite{Anashin:2015woa} (see figure~\ref{fig:Rofs}).
A big step in improving low energy cross section measurements has been
possible with the radiative return or Initial State Radiation (ISR)
method figure~\ref{fig:ISRvsScan} which has been pioneered by the KLOE Collaboration, followed by
BaBar and BES3 experiments.
Recent new experimental input for HVP has been obtained by CMD-3 and
SND at VEPP-2000 via energy scan and by BESIII at PEPC in the ISR
setup (see Contributions by G.~Venanzoni, S.~Eidelman and A.~Denig).
\begin{figure}
\centering
\includegraphics[height=5cm]{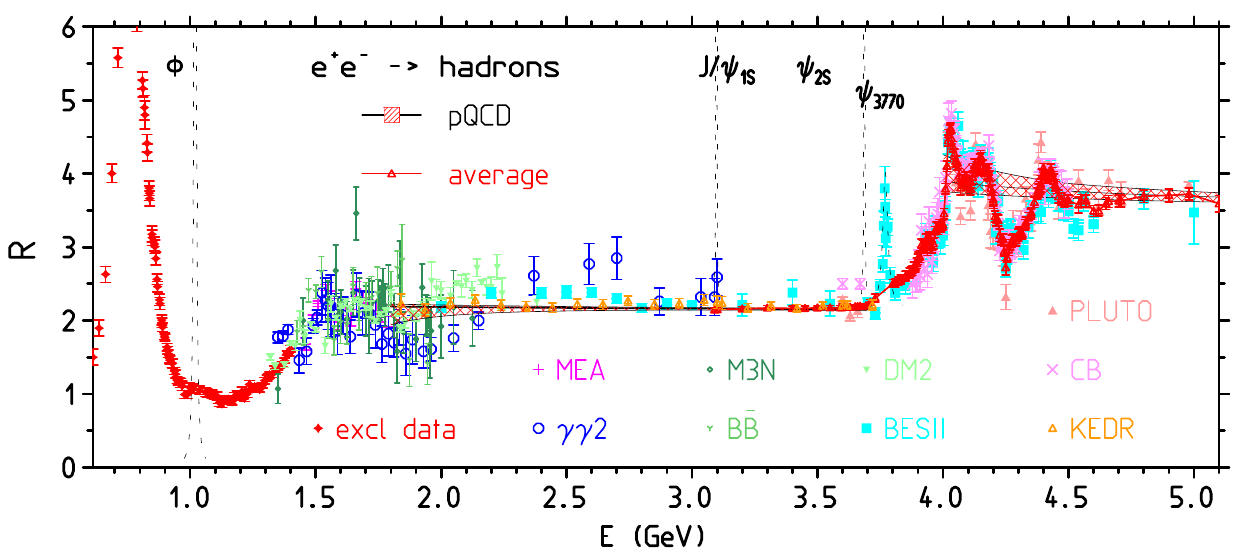}
\includegraphics[height=5cm]{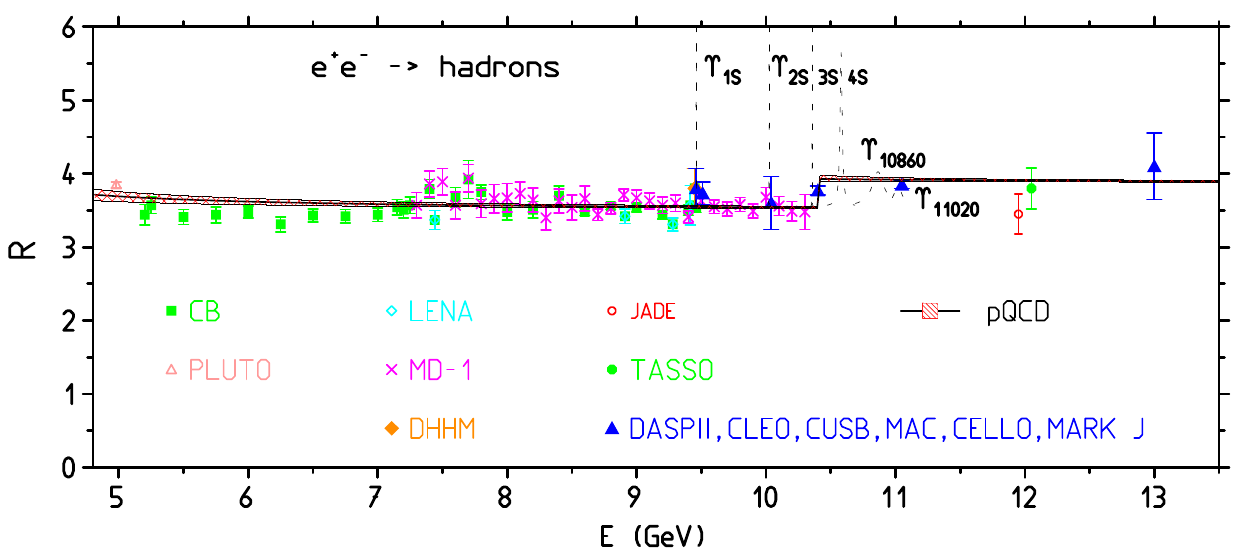}
\caption{Compilation of the $R(s)$ data. New are the KEDR data between
1.84 and 3.72 GeV. Perturbative QCD predictions are also shown.}
\label{fig:Rofs}
\end{figure}
\begin{figure}
\vspace*{-0.7cm}
\centering
\includegraphics[height=3.1cm]{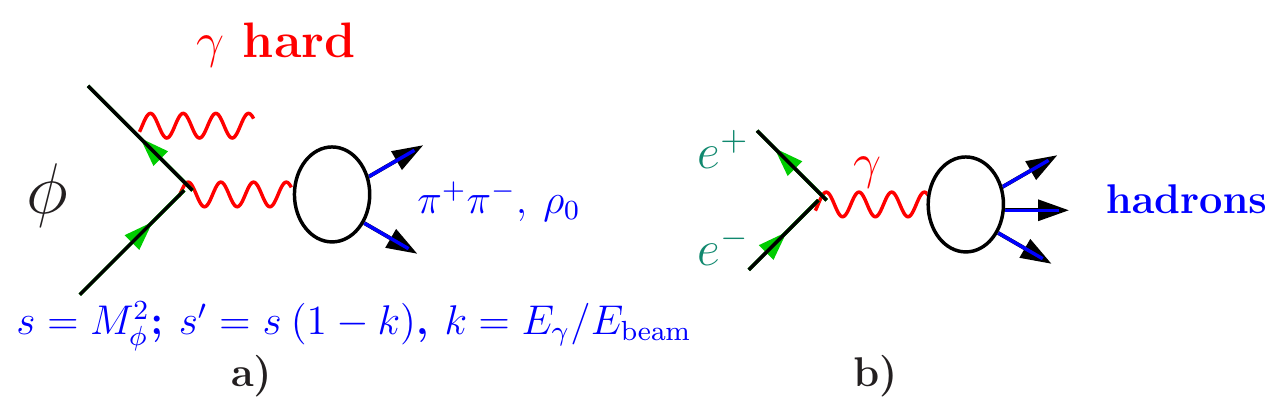}
\caption{ a) Initial state radiation (ISR) on resonance,~~ b) Standard energy scan
by tuning the beam energy.}
\label{fig:ISRvsScan}
\end{figure}

\section{NLO and NNLO HVP effects}
\label{sec-2}
\begin{figure}
\vspace*{-1.0cm}
\centering
\includegraphics[width=7cm]{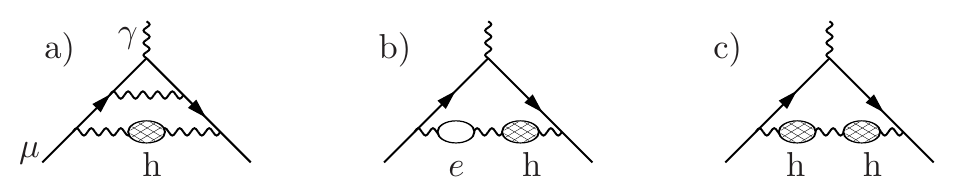}\\[3mm]
\includegraphics[width=9cm]{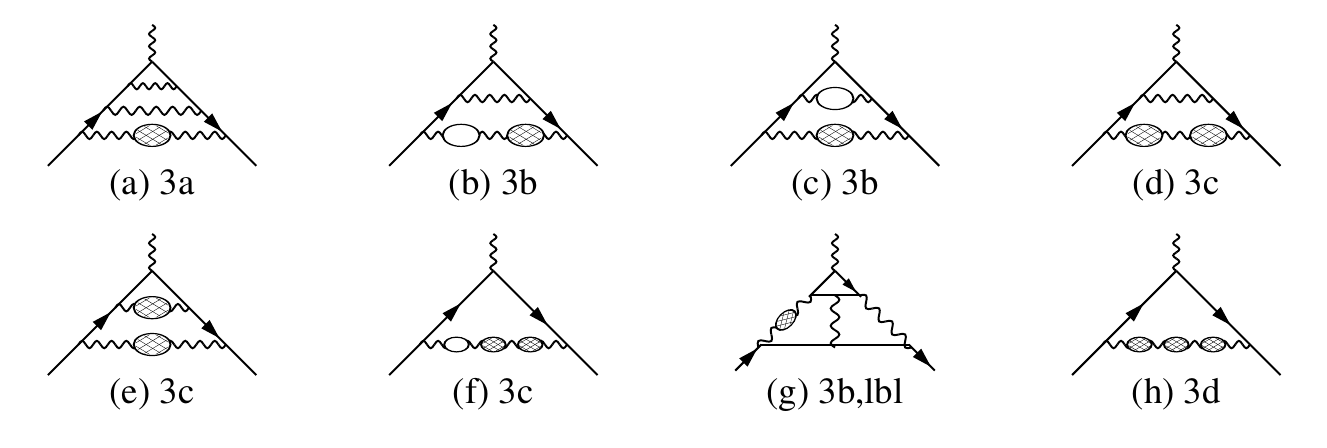}
\caption{Feynman diagrams with hadronic insertions at NLO (top row) and NNLO.}
\label{fig:ammhohad}
\end{figure}

The next-to-leading order (NLO) HVP is represented by diagrams in
figure~\ref{fig:ammhohad}.
With kernels from~\cite{NLO}, the results of an updated evaluation are
presented in table~\ref{tab:amuNLO}.
\begin{table}[h]
\centering
\caption{NLO contributions diagrams a) - c) (in units $10^{-11}$)}
\label{tab:amuNLO}
{\small
\begin{tabular}{cccc}
\hline\noalign{\smallskip}
 $a_\mu^{(2a)}$ &$a_\mu^{(2b)}$ &$a_\mu^{(2c)}$ &$a_\mu^{{\rm had}(2)}$ \\
\noalign{\smallskip}\hline\noalign{\smallskip}
 -206.13(1.30) & 103.49(0.63)& 3.37(0.05) & -99.27 (0.67) \\ 
\noalign{\smallskip}\hline
\end{tabular}}
\caption{NNLO contributions diagrams (a) - (h) (in units $10^{-11}$)}
\label{tab:amuNNLO}
{\small
\begin{tabular}{llllllc}
\hline\noalign{\smallskip}
 $a_\mu^{(3a)}$ &
 $a_\mu^{(3b)}$ &
 $a_\mu^{(3b,\mathrm{lbl})}$ &
 $a_\mu^{(3c)}$ &
 $a_\mu^{(3d)}$ & $a_\mu^{\mathrm{had}(3)}$ & Ref.\\
 \noalign{\smallskip}\hline\noalign{\smallskip}
 $\bl 8.0$        &
 $-4.1$     	  &
 $\bl 9.1$  	  &
 $-0.6$     	  &
 $\bl 0.005$  &  $12.4(1)$ & ~\cite{NNLO}    \\
 $\bl 7.834~(61)$   &
 $-4.033~(28)$ 	  &
 $ \bl 9.005~(63)$  &
 $-0.569~(5)$	  &
 $\bl 0.00518~(12)$ &  $12.24~(10)$ & ~\cite{Jegerlehner:2015stw}\\
\hline\noalign{\smallskip}
\end{tabular}}
\end{table}
The next-to-next leading order (NNLO) contributions have been
calculated recently~\cite{NNLO}. Diagrams are shown in
figure~\ref{fig:ammhohad} and corresponding contributions evaluated
with kernels from~\cite{NNLO} are listed in table~\ref{tab:amuNNLO}.

\section{News on VP subtraction}
\begin{figure}[h]
\vspace*{-4mm}
\centering
\includegraphics[width=0.46\textwidth]{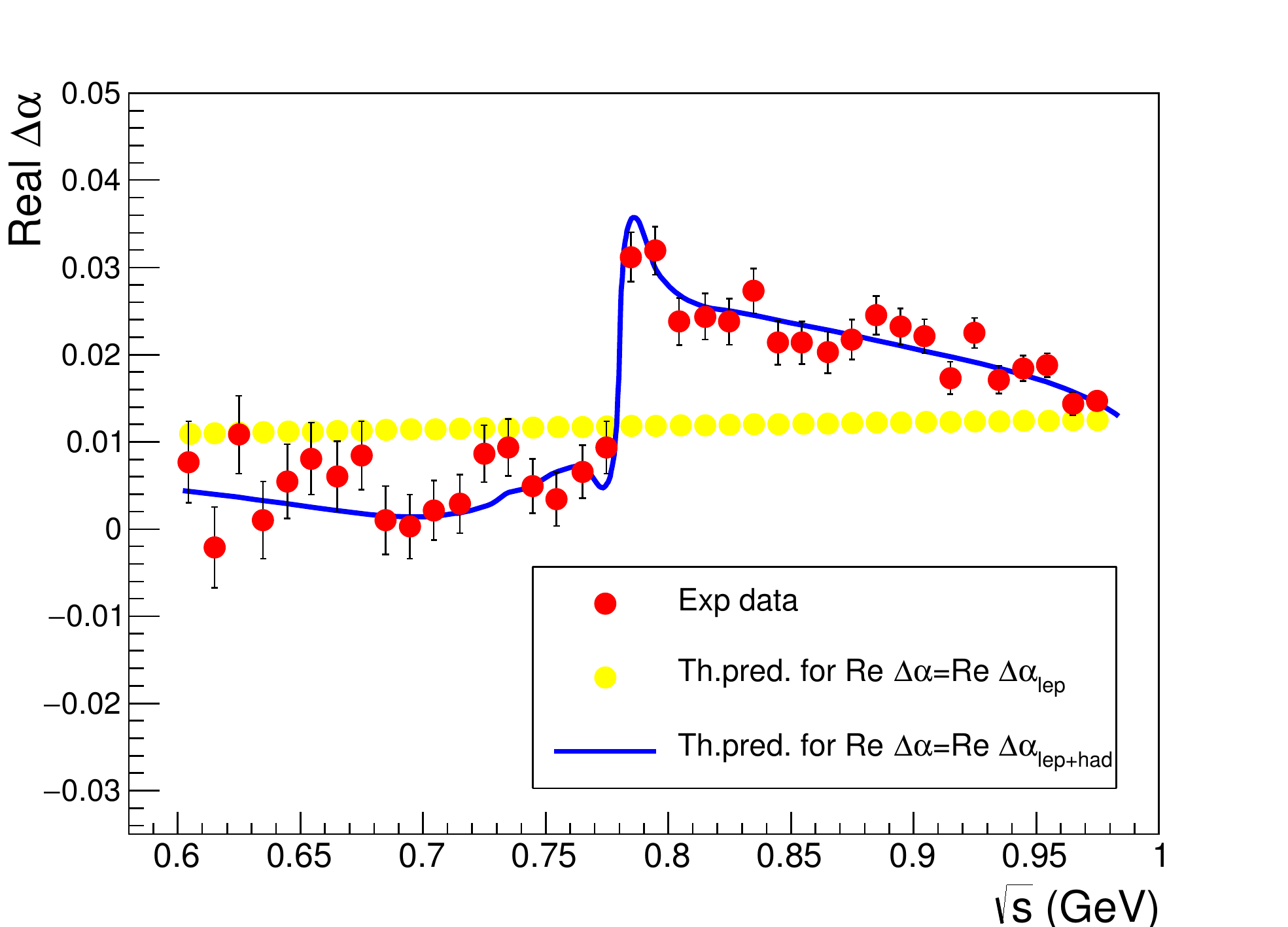}
\includegraphics[width=0.46\textwidth]{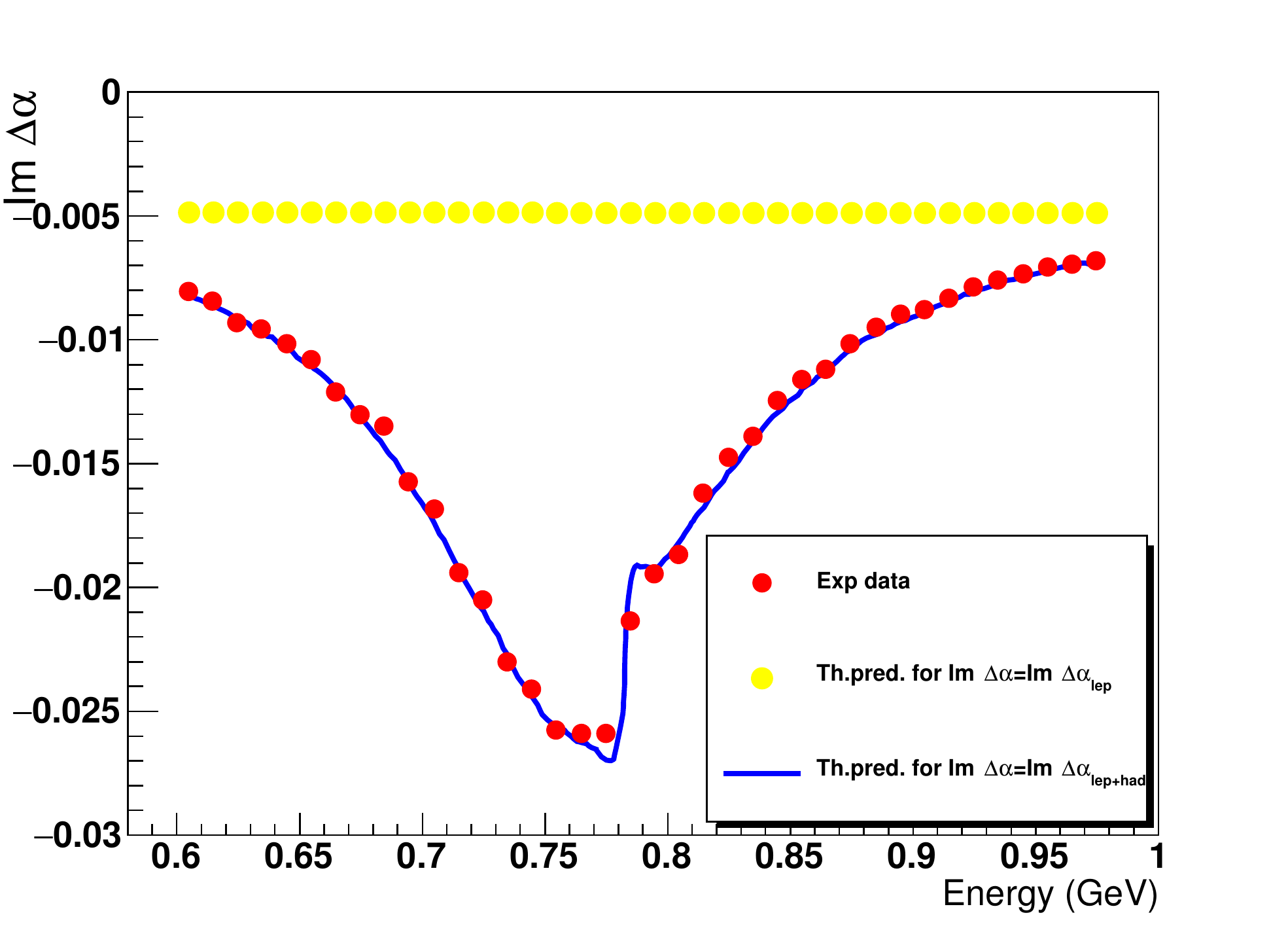}
\caption{Real and imaginary part of the complex shift $\Delta \alpha$
of the fine structure constant as measured by KLOE. Reprinted from~\cite{Anastasi2016bvn}
in Phys.\ Lett. B {\tt doi:10.1016/j.physletb.2016.12.016}}
\label{fig:dalpKLOE}
\end{figure}
The first direct measurement of the timelike complex VP function in the $\rho$
resonance region by KLOE~\cite{Anastasi2016bvn} (see
figure~\ref{fig:dalpKLOE}) nicely confirms
dispersion relation calculation and demonstrates the importance of
including the imaginary part in vacuum polarization subtraction in
obtaining the undressed $\sigma^{(0)}(s)$ version of the physical
hadronic cross-sections $\sigma(s)$. The
complex running fine structure constant
$\alpha(s)=\frac{\alpha(0)}{1-\Delta \alpha(s)}$ is defined in terms
of the complex shift $\Delta \alpha(s)=-[\Pi'_\gamma(s)-\Pi'_\gamma(0)]$.
Measuring $\left|\frac{\alpha(s)}{\alpha(0)}\right|^2=\frac{\sigma(\epm \to \mu^+\mu^-)}
{\sigma(\epm \to \mu^+\mu^-)_{\rm pt}}$
as well as
$R(s)=\frac{\sigma(\epm \to \pi^+\pi^-)}{\sigma(\epm \to
\mu^+\mu^-)}$, which determines $\Impa\,
\alpha(s)=-\frac{\alpha}{3}\,R(s)$, and knowing the modulus $|\alpha(s)|$ one
can extract $\Repa\, \alpha(s)$ as well (see G.~Venanzoni's
Contribution fort details).

The imaginary parts in the perturbative regions usually are small
relative to the leading logarithms which govern the running couplings
(renormalization group approach). In the hadronic shift however,
resonances are accompanied by imaginary parts which may be huge
in particular near resonances which can decay via OZI suppressed
channels only (see Sect.~5 of ~\cite{Jegerlehner:2015stw}).

\section{Low energy effective Lagrangian theory}
\label{sec-4}
\begin{figure}[h]
\vspace*{-9mm}
\centering
\includegraphics[width=0.9\textwidth]{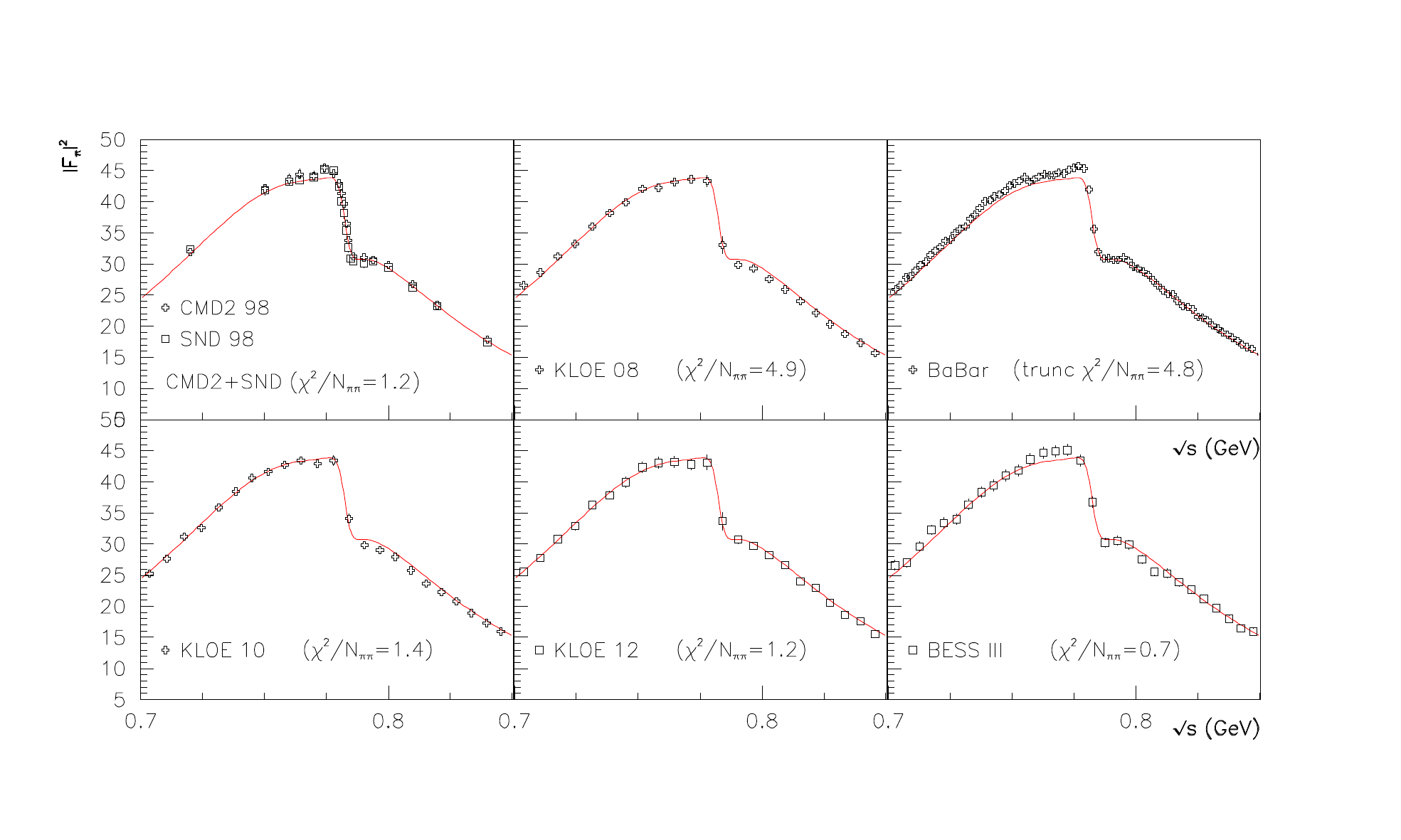}
\caption{Comparing the $\tau$+PDG prediction (red curve) of the pion form factor
in $e^+ e^-$ annihilation in the $\rho-\omega$ interference
region. Reprinted from~\cite{Benayoun:2011mm}, with kind permission of
The European Physical Journal (EPJ).}
\label{fig:tautoepem}
\end{figure}
\begin{figure}
\centering
\includegraphics[height=7cm]{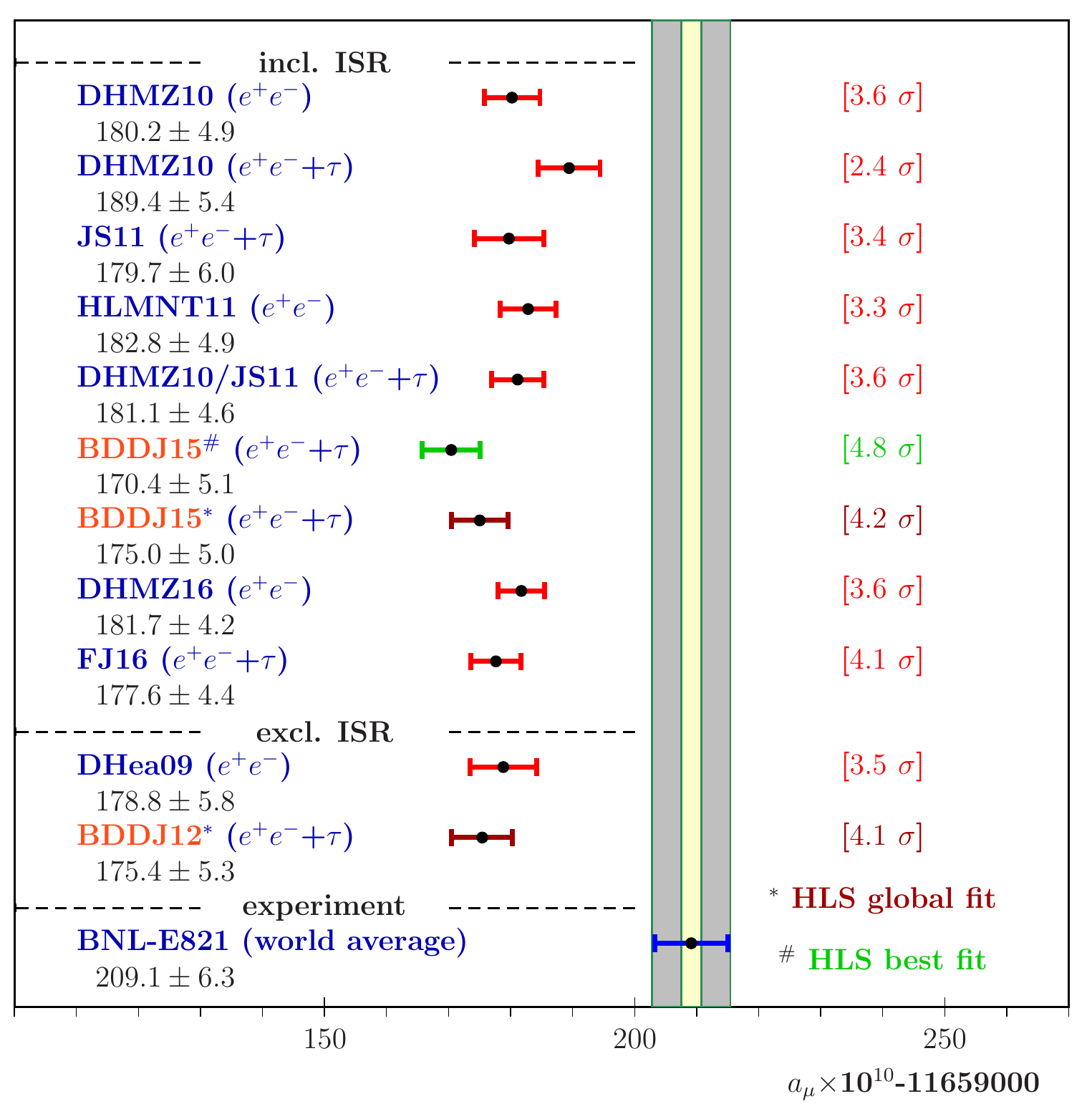}
\caption{Dependence of $a_\mu$ predictions on recent evaluations of
$a_\mu^{\rm had,LO}$. The HLS best fit BDDJ15{$^{\#}$}
(NSK+KLOE10+KLOE12) does not include BaBar $\pi\pi$
data~\cite{Benayoun:2015gxa}, while BDDJ15{$^*$} does.
JS11~\cite{JS11}, FJ16~\cite{Jegerlehner:2015stw} is updated and includes the BES III
and KEDR data. Further points are BDDJ12~\cite{Benayoun:2011mm},
DHMZ10~\cite{Davier:2010nc}, DHMZ16~\cite{Davier:2016udg,Davier:2016iru}, HLMNT11~\cite{HMNT11} and
DHea09~\cite{Davier:2009ag}.
The DHMZ10 ($\epm$+$\tau$) result is not including the $\rho-\gamma$
mixing correction, i.e. it misses important isospin breaking
effects. In contrast, DHMZ10/JS11 is obtained by including this
correction, which brings the point into much better agreement with standard
analyses based on $\epm$ data alone, as for example the DHMZ10 ($\epm$) result.
(see also~\cite{Davier:2015bka,Zhang:2015yfi}). Note: results depend on which
value has been taken for HLbL. JS11 and BDDJ13 includes
$116(39)\power{-11}$ [JN]~\cite{JN},
DHea09, DHMZ10, HLMNT11 and BDDJ12 use $105(26)\power{-11}$
[PdRV]~\cite{PdRV}, while FJ16 includes an updated $103(29)\power{-11}$.}
\label{fig:compare}
\end{figure}

A low energy effective field theory description of hadronic data
requires an extension of CHPT towards higher energies, which mainly
requires to include spin 1 resonances \mbo{\rho,\omega,\phi,\cdots} in
accord with the symmetries of QCD. Principles to be included are the
chiral structure of QCD, VMD and electromagnetic gauge invariance. The
spin 1 resonances are important in the evaluation of HVP as well as
of HLbL effects. One possible implementation is the HLS model, which for
what concerns HVP can also be seen as a generalized Gounaris-Sakurai
model. In the neutral channel $\epm \to
\mathrm{hadrons}$
\mbo{\gamma,\rho^0,\omega,\phi} mixing makes the channel rather complicated, while in the
charged channel of $\tau \to \nu_\tau
\mathrm{hadrons}$~\cite{ALEPH,AlephCorr,OPAL,CLEO,Belle} is much simpler as
the $\rho^\pm$ do not mix with other hadrons. It is thus tempting to
start with the isospin rotated $\tau^\pm\to \nu_\tau \pi^\pm\pi^0$ decay
spectra and supplement them with appropriate isospin breaking and
mixing effects to predict $\epm \to\pi^+\pi^-$, with the result shown in
figure~\ref{fig:tautoepem}. It shows that there is no $\tau$ vs. $\epem$
conflict and actually simultaneous fits allows one to reduce uncertainties of
HVP by using indirect constraints. The global fit strategy followed
in~\cite{Benayoun:2011mm} takes into account data below
\mbo{E_0 = 1.05\gv} (just above the
\mbo{\phi}) to constrain the effective Lagrangian couplings. Used are 45
different data sets, 6 annihilation channels and 10 partial width
decays. The effective theory then allows us to predict cross sections
for the channels
\mbo{\ppm,\pi^0\gamma,\eta\gamma,\eta'\gamma,\pi^0\pi^+\pi^-,K^+K^-,K^0\bar{K}^0\,,}
which account for 83.4\% of $\amuh$. The missing channels
\mbo{4\pi,5\pi,6\pi,\eta\pi\pi,\omega\pi} and the higher energy tail
\mbo{E>E_0}
is evaluated using data directly and pQCD for the perturbative region and tail.
All mixing effects, as \mbo{\gamma\rho}-mixing,
\mbo{\rho\omega}-mixing, $\cdots$, as well as the decay branching fractions are
dynamically generated by including self-energy effects of the spin 1
mesons. One thus is taking into account proper phase space, energy dependent widths
etc. Such fit strategy is able to shed light on incompatibilities in
the data, e.g. KLOE vs BaBar, by comparing the fit qualities, but also
reveals the compatibility of $\tau$--decay spectra with $\epm$--data
after accounting for the mixing effects like including $\gamma-\rho^0$
mixing. HLS estimates are included in table~\ref{fig:compare} together
with other recent results.

\section{HVP from lattice QCD (following H. Wittig at LATTICE 2016)}
\label{sec-5}
The need for ab initio calculation of $a_\mu^{\rm had}$ is well
motivated: -- the problems to determine non-perturbative contributions
to the muon $g-2$ from experimental data at sufficient precision
persists and is not easy to improve, -- a model--independent extension
of CHPT to the relevant energies ranges up to 2 GeV is missing, while
the new experiments E989 \@ FNAL and E34 \@ J-PARC require an
improvement of the hadronic uncertainties by a factor two to four.

The hope is that LQCD can deliver estimates of accuracy
\bea
\delta a_\mu^{\rm HVP}/a_\mu^{\rm HVP} < 0.5 \% \;,\;\;\delta a_\mu^{\rm
HLbL}/a_\mu^{\rm HLbL} \lapprox  10 \%
\eea
in the coming years.

Primary object for getting HVP in LQCD is the e.m. current correlator in configuration
space
\bea
\langle J_\mu(\vec{x},t)\,J_\nu(\vec{0},0)\rangle\semis
J_\mu=\frac23 \bar{u}\gamma_\mu u-\frac13 \bar{d}\gamma_\mu
d-\frac13 \bar{s}\gamma_\mu s + \cdots
\eea
In principle, a Fourier transform
\bea
\Pi_{\mu\nu}(Q)=\int \D^4 x \E^{\I\,Q x}\,\langle
J_\mu(x)\,J_\nu(0)\rangle=\left(Q_\mu Q_\nu-\delta _{\mu\nu}\,Q^2\right)\,\Pi(Q^2)
\eea
yields the vacuum polarization function $\Pi(Q^2)$ needed to calculate
\bea
a_\mu^{\rm HVP}=4\alpha^2\int_0^\infty \D Q^2\, f(Q^2)\,\left\{\Pi(Q^2)-\Pi(0)\right\}\epo
\label{amuintLQCD}
\eea
The integration kernel in this representation is
\bea
f(Q^2)=w(Q^2/m_\mu^2)/Q^2\semis w(r)=\frac{16}{r^2 \left(1+\sqrt{1+4/r}\right)^4\,\sqrt{1+4/r}}\epo
\eea
\begin{figure}[h]
\centering
\includegraphics[width=0.5\textwidth]{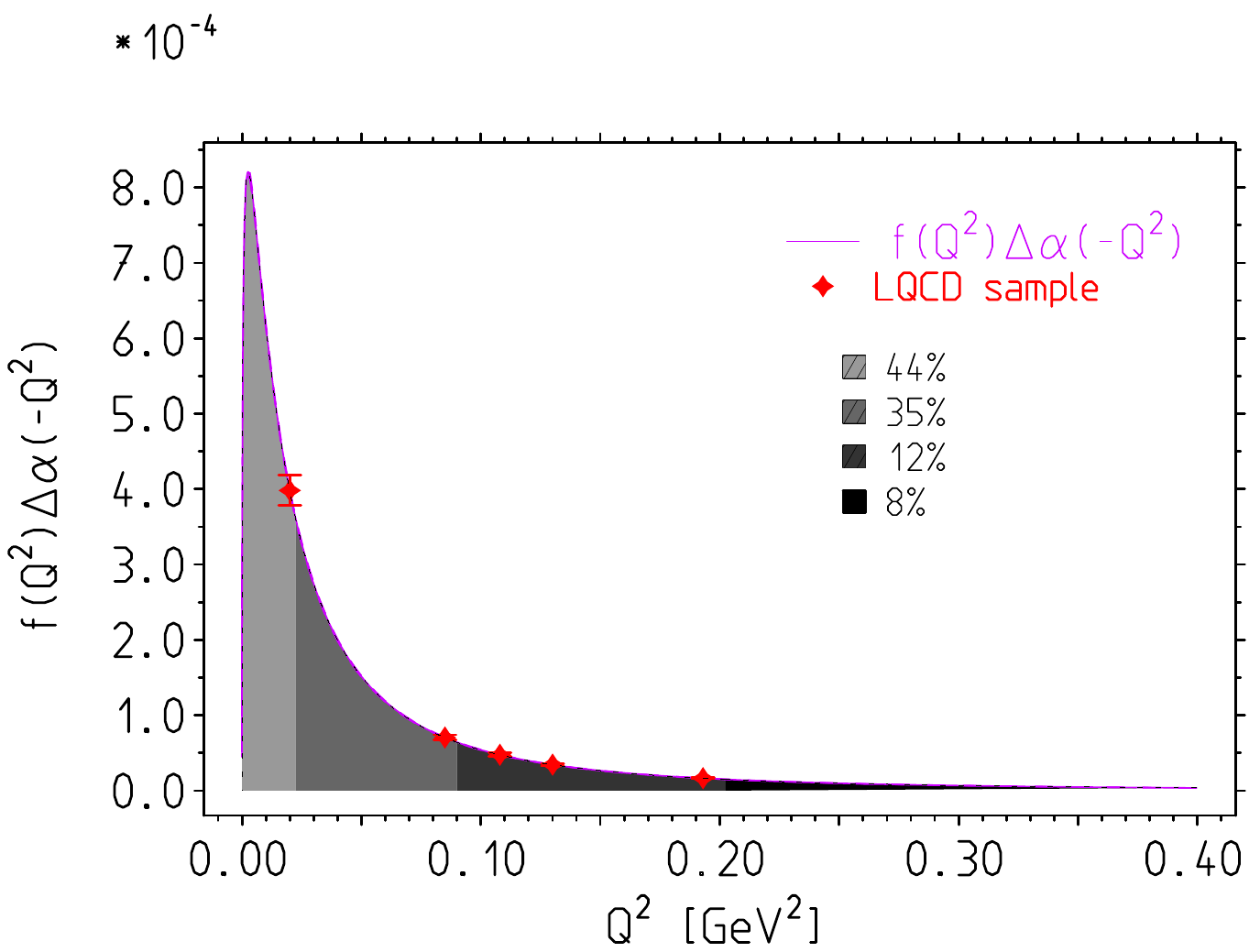}
\includegraphics[height=5cm]{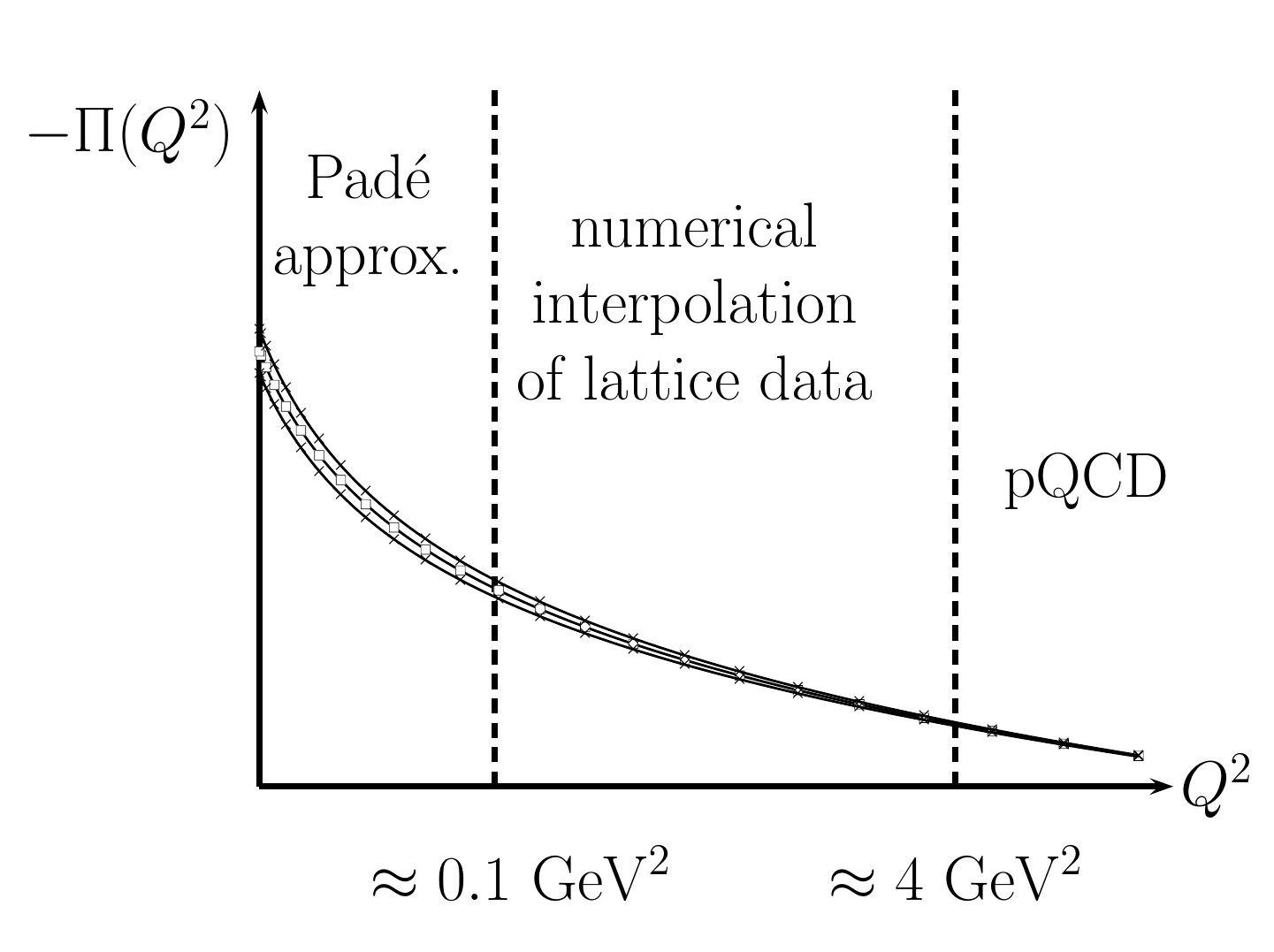}\\
\caption{Left: the integrand of (\ref{amuintLQCD}) as a function of
$Q^2$. Also shown are ranges between
$Q_i=0.00,\,0.15,\,0.30,\,0.45$ and $1.0\gv$
and their percent contribution to $\amuh$ and a possible ``LQCD
sample''. Right: typical ranges for lattice data and their extrapolation to low $Q^2$.}
\label{fig:amuintLQCD}
\end{figure}
As LQCD per se has to work on a lattice in a finite box, momenta are
quantized $Q_{\rm min}=2\pi/L$, where $L$ is the lattice box
length. Therefore, approaching low momenta $Q_{\rm min} \to 0
\Leftrightarrow L\to \infty$ requires a sufficiently large volume.
Present state of the art calculations reach $Q_{\rm min}=2\pi/L$ with
$m_\pi aL\gapprox 4$ for $ m_\pi\sim 200\mv$, such that $ Q_{\rm
min}\sim 314\mv$. This means that about 44\% of the low $Q$ contribution to
$\amuh$ is not covered by data yet.
Typically, lattice data are available for $ Q^2>
\left(2\pi/L\right)^2$, which one has to
extrapolate to $ Q^2=0$ by VMD type modeling~\cite{Feng:2011zk} or
via Pad\'e's~\cite{Golterman:2014ksa} or analytic
continuation~\cite{Feng:2013xsa}. The method requires a reliable
estimate of the bare $\Pi(0)$ (see e.g. ~\cite{Borsanyi:2016lpl}). In order to reach the required accuracy
one needs LQCD data down to $Q^2_{\rm min}\approx 0.1\gv^2$.
\cite{Boyle:2011hu,Feng:2013xsa,Aubin:2013daa,Francis:2014dta,Malak:2015sla}
\begin{figure}[h]
\vspace*{-6mm}
\centering
\includegraphics[width=0.60\textwidth]{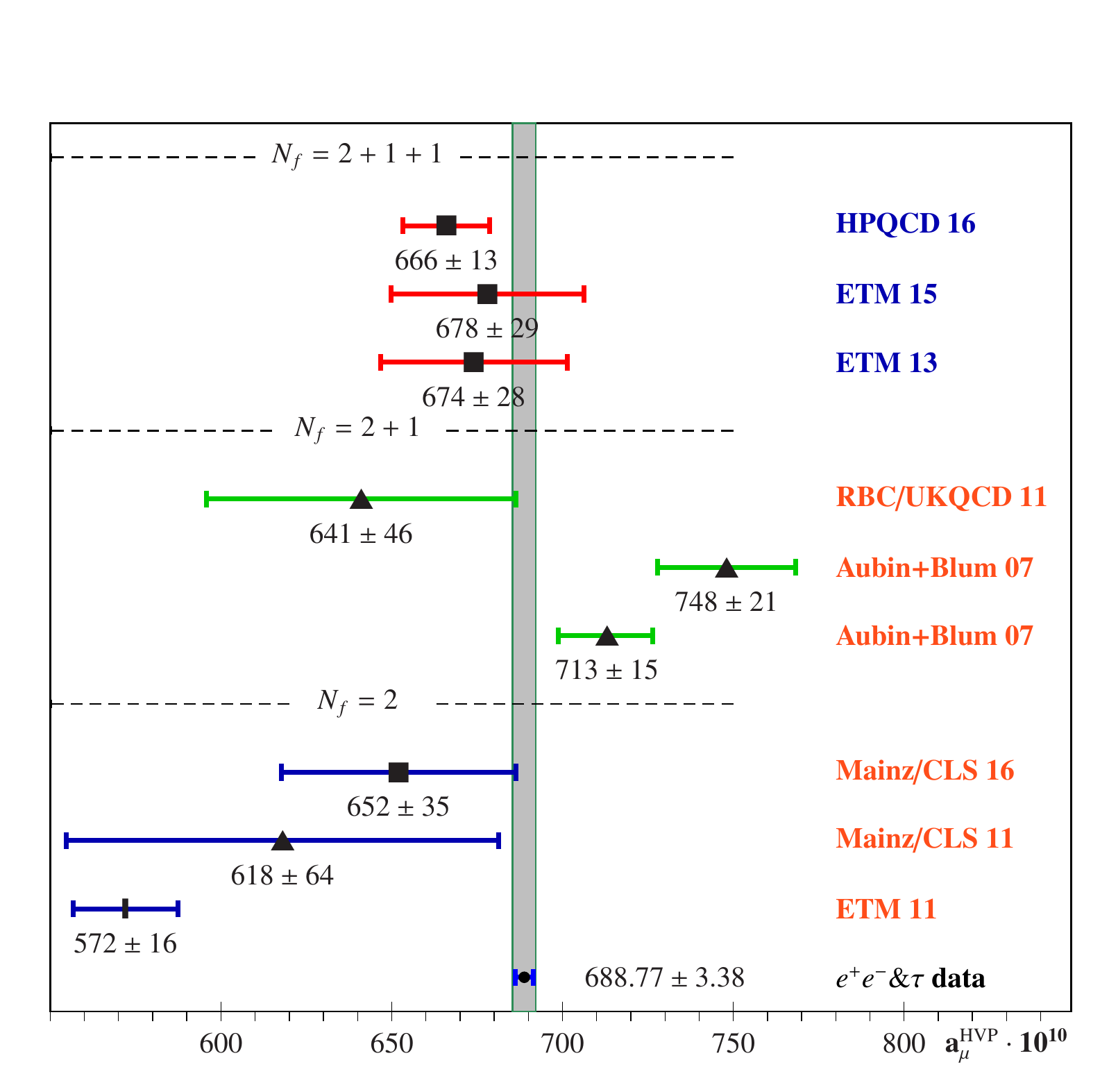}
\caption{Summary of recent LQCD results for the leading order
$a_\mu^{\rm HVP}$, in units $10^{-10}$. Labels: \ding{110} marks
$u,d,s,c$, \ding{115} $u,d,s$ and \ding{121} $u,d$
contributions. Individual flavor  contributions from light $(u,d)$
amount to about 90\%, strange about 8\% and charm about 2\%. Results
shown are from  HPQCD~16~\cite{Chakraborty:2016mwy}, ETM~15~\cite{Burger:2015hdi},
ETM~13~\cite{Burger:2013jya}, RBC/UKQCD~11 ~\cite{Boyle:2011hu},
Aubin+Blum~07~\cite{Aubin:2006xv}, Mainz/CLS~16~\cite{DellaMorte:2016izp},
Mainz/CLS~11~\cite{DellaMorte:2011aa} and
ETM~11~\cite{Feng:2011zk}. The vertical band shows the $\epm$ data
driven DR estimate (\ref{LOHVP}).}
\label{fig:gm2LQCDHVP}
\end{figure}
Some recent results are collected in figure~\ref{fig:gm2LQCDHVP}.
\section{Alternative method to get $\amuh$: using $\alpha(t=-Q^2)$
measured via $t$--channel exchange processes.}
\label{sec-6}
\begin{figure}[h]
\vspace*{-9mm}
\centering
\includegraphics[width=0.35\textwidth]{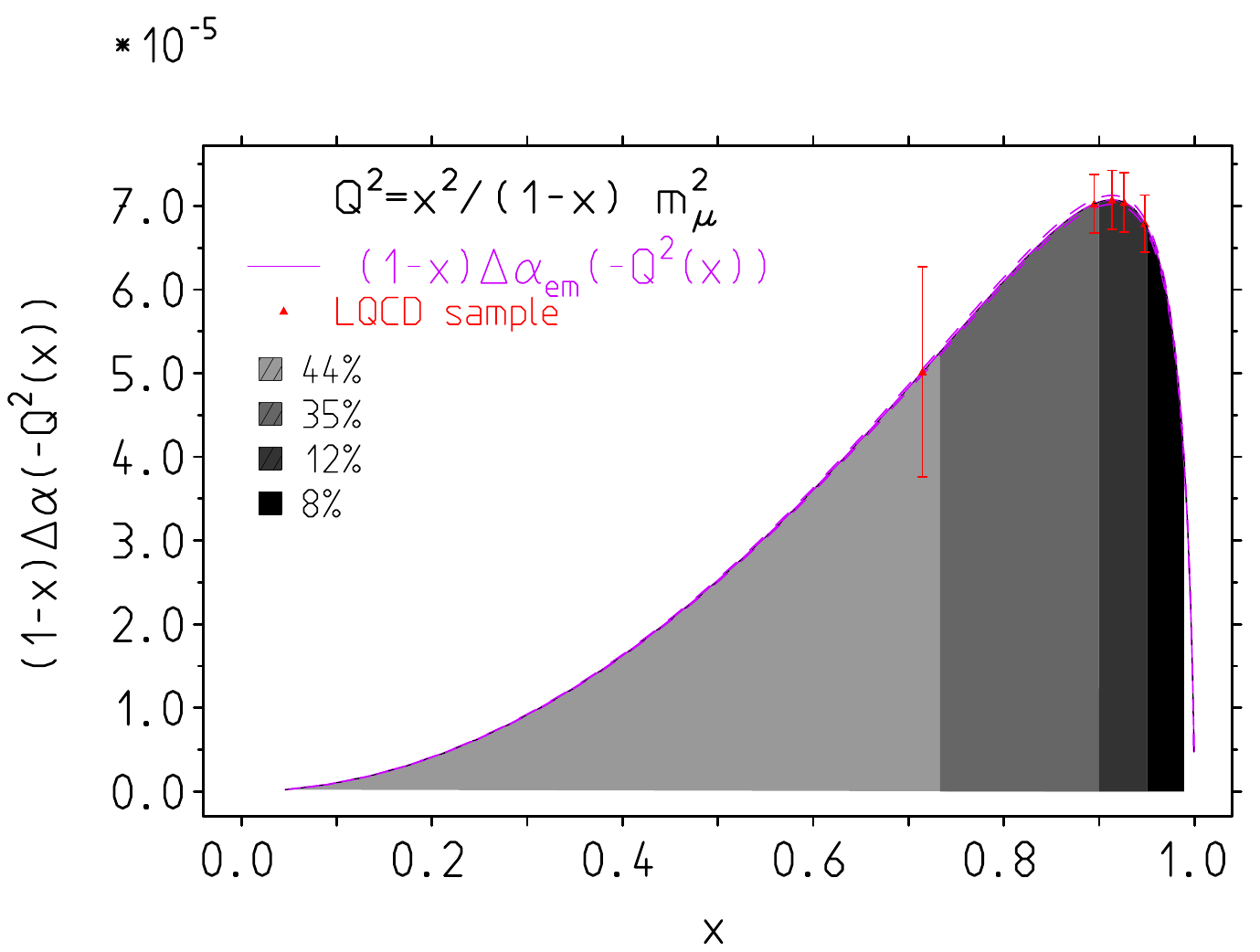}
\includegraphics[width=0.35\textwidth]{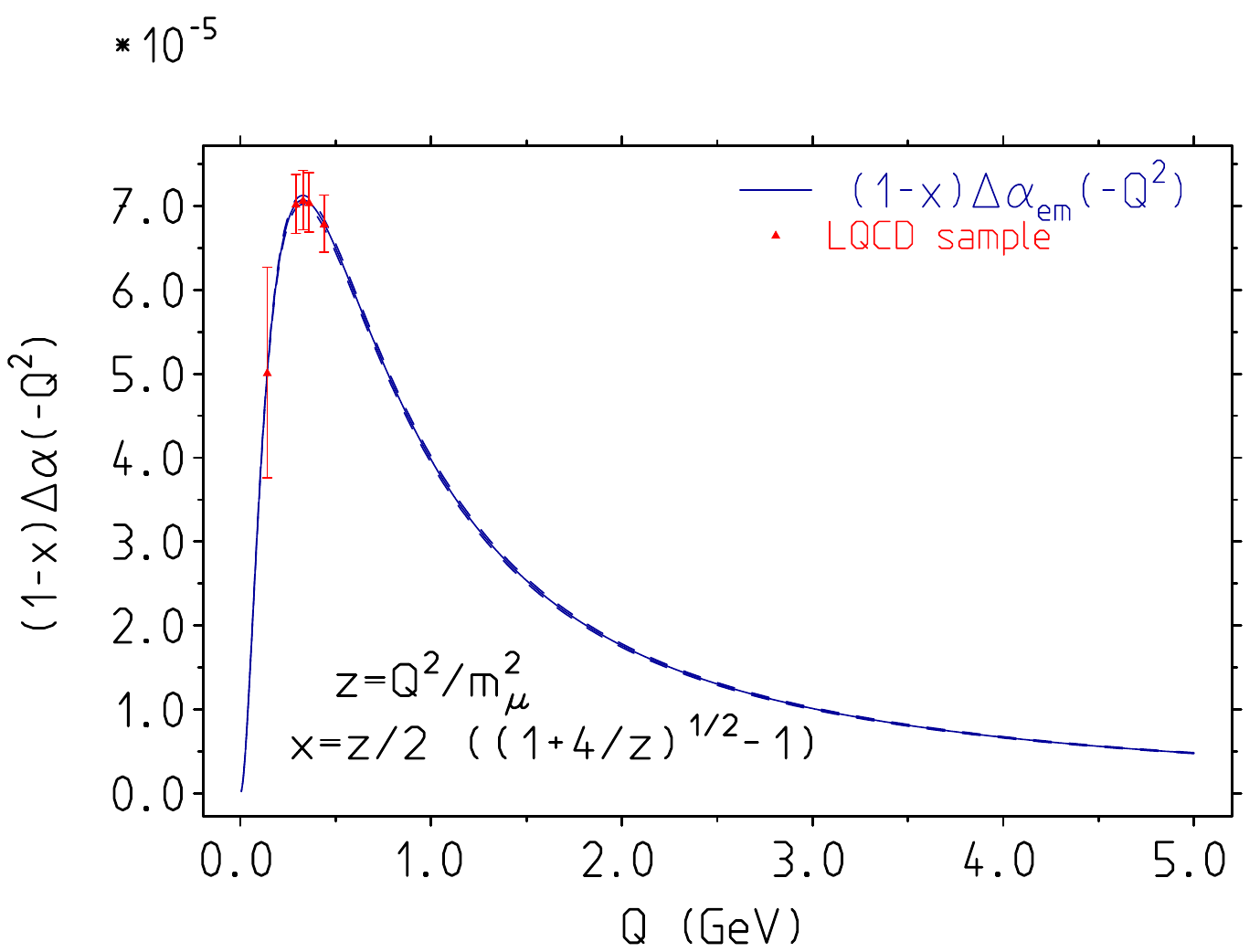}
\caption{The integrand of the \mbo{\amuh} integral (\ref{amuintofx})
as functions of $x$ and $Q$ is strongly
peaked at about $ 330\mv$. Legend as in figure~\ref{fig:amuintLQCD}.}
\label{fig:RAIkernels}
\end{figure}

A promising alternative method to determine $\amuh$ is possible by a
dedicated measurement of $\alpha(t)$ at spacelike
momentum transfer as advocated in~\cite{Calame:2015fva}
and~\cite{Abbiendi:2016xup}. Given $\alpha(-Q^2)$ and the fact that
the leptonic contribution is well under control in perturbation theory
one can extract the hadronic shift
\bea
\dalh(-Q^2)=1-\frac{\alpha}{\alpha(-Q^2)}-{\Delta \alpha^{\rm lep}(-Q^2)}
\eea
and determine \mbo{\amuh} via the representation
\bea
\amuh=\frac{\alpha}{\pi}\int\limits_0^1 \D x\:(1-x)\: \dalh
\left(-Q^2(x)\right)
\label{amuintofx}
\eea
where $ Q^2(x)\equiv \frac{x^2}{1-x}m_\mu^2$ is the spacelike
square momentum--transfer. In the Euclidean region the integrand is highly peaked
around half of the $\rho$ meson mass scale (see figure~\ref{fig:RAIkernels}).
The method is very different from the standard approach based on
(\ref{amuDRbasic}): radiative corrections are very different (much
simpler) as no hadronic final states need to be understood, no VP
subtraction is to be performed, no exclusive channel collection etc.
So, even a 1\% level measurement can provide important independent
information. This in view of the problem to get accurate hadronic
total cross--section in the range between 1 and 2 GeV and possible
unsettled problems (non-convergence of the Dyson resummation near OZI
suppressed resonances) in VP subtraction as addressed recently in
Sect.~5 of~\cite{Jegerlehner:2015stw}.

\begin{figure}[h]
\centering
\includegraphics[height=3cm]{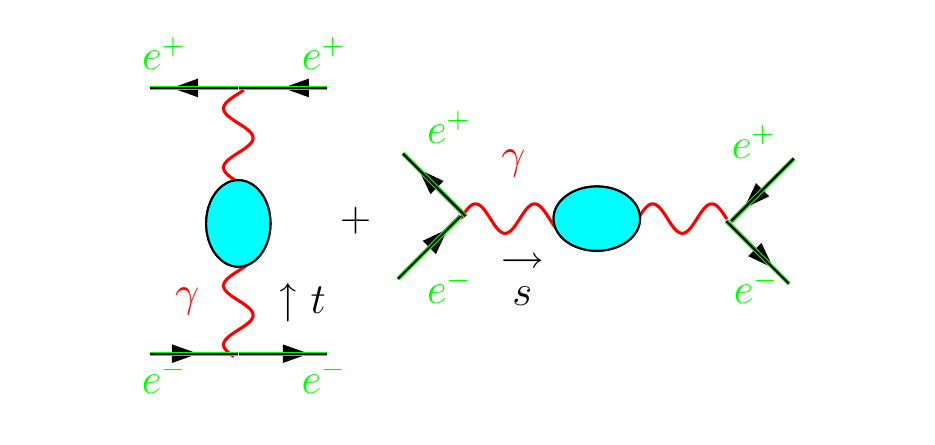}
\includegraphics[height=3cm]{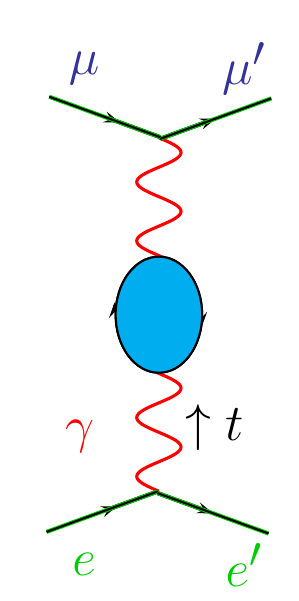}
\caption{Measuring the running charge in the spacelike regime.
Left: VP dressed tree level Bhabha scattering in QED, for small $t$
(small angels) the $s$ channel is suppressed.
Right: getting $\alpha(t)$ from the $\mu^-e^- \to \mu^-e^-$ process.}
\label{fig:processes}
\end{figure}

The possible processes to measure $\alpha(t)$ are Bhabha
scattering $ e^+(p_+)\:\:e^-(p_-) \to e^+(p'_+)\:\:e^-(p'_-)$
or muon electron scattering (see figure~\ref{fig:processes}).
The Bhabha process has two tree level diagrams a $t$-- and a
$s$--channel one. With the
positive c.m. energy square $ s=(p^++p^-)^2$ and the negative momentum
transfer square $ t=(p_--p'_-)^2=-\ha \: (s-4m_e^2)\:(1-\cos
\theta)\,,$ $ \theta$ the $ e^-$ scattering angle, there are two very
different scales involved which helps to isolate the $t$ channel of interest.
The VP dressed lowest order cross--section is
\bea
\frac{\D \sigma}{\D \cos \Theta}= \frac{s}{48\pi}\:\sum\nolimits_{ik}|A_{ik}|^2\,,
\eea
where $ A_{ik}$ are tree level helicity amplitudes, $i,k=$L,R denote left--
and right--handed electrons.

The dressed transition amplitudes in the massless limit
(\mbo{m_e\approx 0}) read
\bea
|A_{\rm LL,RR}|^2=\frac{3}{8}\:(1+\cos
\theta)^2\:\left|\frac{e^2(s)}{s}+\frac{e^2(t)}{t}\right|^2 \semis
|A_{\rm LR,RL}|^2=\frac{3}{8}\:(1-\cos
\theta)^2\:\left|\frac{e^2(s)}{s}+\frac{e^2(t)}{t}\right|^2 \epo
\eea
Preferably one uses small angle Bhabha scattering (small $|t|$) as a
normalizing process which is dominated by the $t$--channel $\sim
1/t$, however, detecting electrons and positrons along the beam axis
often has its technical limitations. Care also is needed concerning
the ISR corrections because cuts for the Bhabha process ($ \epm \to
\epm$) typically are different from the ones applied to { $\epm \to$
hadrons}. Usually, experiments have included corresponding
uncertainties in their systematic errors, if they have not been explicitly
accounted for by applying appropriate radiative corrections. For details I
refer to~\cite{Calame:2015fva} and the Contribution by Luca Trentadue.

While the Bhabha process requires to sort out the $s$ channel from
the $t$ channel, the pure $t$ channel reaction of $\mu^- e^-$ scattering $ \mu^-(p_-)\:\:e^-(q_-)
\to \mu^-(p'_-)\:\:e^-(q'_-)$ provides a much simpler setup and could be
realized as a fixed target
experiment~\cite{Abbiendi:2016xup} at existing
facilities. The leading order cross--section in this case has the
simple form
\bea
\frac{\D \sigma^{\rm unpol.}_{\mu^-e^- \to \mu^-e^-}}{\D
t}=4\pi\,{ \alpha(t)^2}\,\frac{1}{\lambda(s,m_e^2,m_\mu^2) }\,\left\{\frac{\left(s-m_\mu^2-m_e^2\right)^2}{t^2}+\frac{s}{t}+\frac12\right\}\,,
\eea
exhibiting the effective charge as an overall factor. For details see
Luca Trentadue's Contribution. Such an experiment would provide data
for the Euclidean electromagnetic current correlator
$\Pi'_\gamma(Q^2)-\Pi'_\gamma(0)=-\dalh(-Q^2)=\frac{\alpha}{\alpha(-Q^2)}+\Delta
\alpha^{\rm lep}(-Q^2)-1$ and would allow for a directly check of lattice QCD data.
In addition one could determine $\dalh\left(-Q^2\right)$
at $ Q \approx 2.5\gv$ by this method (one single number!) as the
non-perturbative part of \mbo{\dalh
\left(M_Z^2\right)=\dalh\left(-Q^2\right)+\mathrm{pert.}} when evaluated in ``Adler
function'' approach advocated in~\cite{Jegerlehner:2008rs}.

\section{Theory confronts experiment}
\label{sec-7}
\begin{table}[h]
\vspace*{-4mm}
\caption{Standard model theory and experiment comparison [in units $10^{-10}$].}
\label{tab:amucontrubitionsummary}
\centering
\small
\begin{tabular}{lr@{ .}lr@{ .}lc}
&\ttc{}&\ttc{}&\\
\hline
&\ttc{}&\ttc{}&\\
Contribution & \multicolumn{2}{c}{Value} & \multicolumn{2}{c}{Error} & Reference \\
&\ttc{}&\ttc{}&\\
\hline\noalign{\smallskip}
QED incl. 4-loops+5-loops & { 11\,658\,471}&{ 8851} & {
0}&{ 036} & Remiddi, Kinoshita et al.\\
Leading hadronic vac. pol.& { 688}&{ 77} & {
3}&{ 38} & data-driven $\epm+\tau$ \\
Subleading hadronic vac. pol. &{ -9}&{ 927 } &
{ 0}&{ 072} & 2016 update  \\
NNLO hadronic vac. pol. &  { 1}&{ 224} &
{ 0}&{ 010} & ~\cite{NNLO}  \\
Hadronic light--by--light &  { 10}&{ 34} & { 2}&{ 88}
& ~\cite{JN,Bijnens:2016hgx} \\
Weak incl. 2-loops & { 15}&{ 36} & { 0}&{ 11} & ~\cite{CMV03,Gnendiger:2013pva}  \\
&\ttc{}&\ttc{}&\\
Theory & { 11\,659\,177}&{ 6} & { 4}&{ 4} & --  \\
Experiment & { 11\,659\,209}&{ 1} & { 6}&{ 3} & ~\cite{BNLfinal} updated  \\
Exp.- The.  {{ 4.1}} standard deviations & { 31}&{ 3}
& { 7}&{ 7} & -- \\ \noalign{\smallskip}\hline
\end{tabular}
\end{table}
Table~\ref{tab:amucontrubitionsummary} summarizes the present status
of the SM prediction for $\amu$ in comparison with the experimental
value~\cite{BNLfinal}. For a recent update of the weak contribution
see~\cite{Gnendiger:2013pva}. As an estimate based
on~\cite{HKS95,BPP1995,KnechtNyffeler01,MV03,JN,Nyffeler:2016xul,Gerardin:2016cqj,Bijnens:2016hgx}
we adopt $\pi^0,\eta,\eta'$ [$95 \pm 12$] + axial--vector [$8 \pm
~3$] + scalar [$-6\pm ~1$] + $\pi,K$ loops [$-20\pm 5$] + quark loops
[$22\pm ~4$] + tensor [$1\pm ~0$] + NLO [$3\pm ~2$] which yields
\bea
 a^{(6)}_\mu(\mathrm{lbl},\mathrm{had})=(103 \pm 29) \power{-11}  \epo
\eea
The result differs little from the ``agreed'' value $(105\pm26)
\power{-11}$ presented in~\cite{PdRV} and $(116\pm39)\power{-11}$
estimated in~\cite{JN}. Both included a wrong, too large, Landau-Yang
theorem violating axial--vector contribution from~\cite{MV03},
correcting for this we obtain our reduced value relative to~\cite{JN}.

\noindent
The following tabular collects recent new/updated evaluations:\\
{ \small
\begin{tabular}{llrcl}
\hline
New contribution & Reference & \multicolumn{3}{c}{\mbo{\Delta \amu\cdot 10^{11}}}\\
\hline
{ NNLO HVP} &\cin{Kurz et al. 2014} & \mbo{12.4}& \mbo{\pm}&\mbo{0.1}\\
{ NLO HLbL} &\cin{Colangelo et al. 2014} & \mbo{3}&\mbo{\pm}&\mbo{2}\\
{ New axial exchange HLbL} & \cin{Pauk, Vanderhaeghen~\cite{PaukVanderhaeghen2013},
FJ14~\cite{FJ14,Jegerlehner:2015stw}} & \mbo{7.55}&\mbo{ \pm}&\mbo{ 2.71}\\
{ Tensor exchange HLbL} & \cin{Pauk, Vanderhaeghen 2014} & \mbo{1.1}&\mbo{ \pm}&\mbo{ 0.1}\\
{ New $\pi^0$ exchange HLbL} & $\pi^0\gamma^*\gamma^*$ constraint from LQCD~\cite{Gerardin:2016cqj}& $64.68$&$\pm$&$12.40$\\
$\cdots$ & & & & $\cdots$\\
{ Old axial exchange HLbL}& \cin{Melnikov, Vainshtein 2004} & $22$&$\pm$&$5$\\
{ Old $\pi^0$ exchange HLbL} &JN~\cite{JN}& $72$&$\pm$&$12$\\
\hline
Total change & & { $-5.6$}&{ $\pm$}& { $12.85$}{ [$\leftarrow 13$]}\\[4mm]
\end{tabular}
}

The uncertainty from these contributions remains unchanged, while the
central value is shifted downwards by almost 1 SD.

Possible interpretations of the 4
$\sigma$ deviation: new physics?, a statistical fluctuation?,
underestimated uncertainties (experimental, theoretical)?  Do
experiments measure what theoreticians calculate?  The challenge for
the future is to keep up with the future experiments, which will
improve the experimental accuracy from $\delta a_\mu^\mathrm{exp} = 63
\times 10^{-11}$ [$\pm$0.54 ppm] at present to $\delta
a_\mu^\mathrm{exp}=16 \times 10^{-11}$ [$\pm$0.14 ppm] the next years.
Next generation experiments require a factor 4 reduction of the
uncertainty optimistically feasible should be a factor 2 we hope.

In view of the upcoming two \textbf{complementary} experiments, one
at Fermilab working with ultra hot muons and the other at J-PARC
operating with ultra cold muons (very different radiation effects),
the big challenge is to keep up on the prediction side as much as
possible. The deviation between theory and experiment can be
scrutinized provided theory and the needed cross section data improves
the same as the muon \mbo{g-2} experiments. Primarily we need
more/better data and/or progress in non-perturbative QCD, where the
main obstacle (data, lattice QCD, RLA) is the hadronic light-by-light
scattering contribution. Progress in evaluating HVP also depends on
more data (BaBar, Belle, VEPP-2000, BESIII,...) and lattice QCD where
recent progress is very promising (see figure~\ref{fig:gm2LQCDHVP}). In
both cases HVP as well as HLbL, lattice QCD will provide answers one
day, but also low energy effective RL and DR approaches need be further
developed. One has also to keep in mind that progress in calculations of radiative
corrections~\cite{Actis:2010gg,CZYZ:2014yra,Jegerlehner:2017kke} is
mandatory in precision measurements of hadronic cross sections.

\begin{figure}[h]
\centering
\includegraphics[height=2cm]{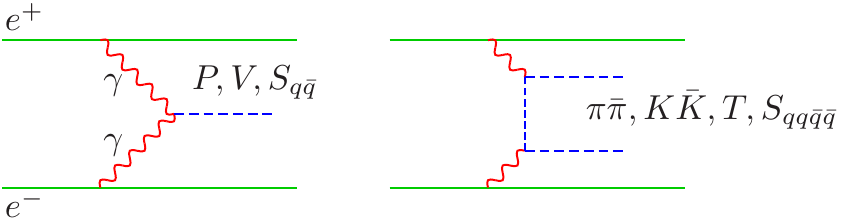}
\includegraphics[height=2cm]{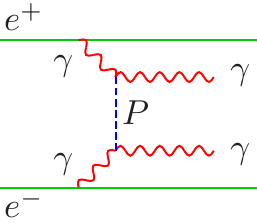}\\
\includegraphics[height=2.00cm]{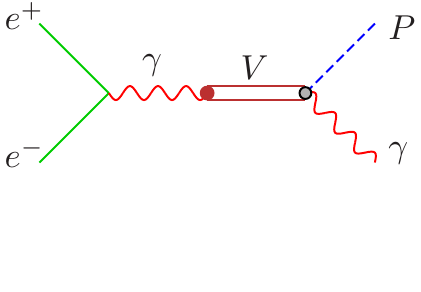}
\includegraphics[height=2.0cm]{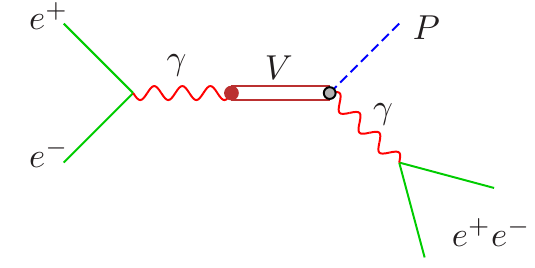}
\caption{Sample processes relevant for the dispersive approach to HLbL.}
\label{fig:gammagammaetal}
\end{figure}

For future improvements of HLbL one urgently needs more information
from $\gamma\gamma \to \mathrm{ \ hadrons \ }$
physics~\cite{Babusci:2011bg,Nyffeler:2013oca} in order to have better
constraints on modeling hadronic amplitudes
(see~\cite{Mennessier:1982fk,Moussallam:2013una,Pennington14} for
theoretical studies). Some sample processes are collected in figure~\ref{fig:gammagammaetal}.
Mostly experiments at $\epm$--facilities investigate single-tag events
(higher rates, lower background). New data are expected from KLOE,
KEDR exhibiting taggers and from BaBar, Belle, BES III which have high
luminosity. More information is also expected from Dalitz--decay
studies $\rho,\omega,\phi \to \pi^0(\eta)e^+e^-$ possible at
Novosibirsk, CERN NA60, JLab, Mainz, Bonn, J\"ulich and BES.
Unfortunately some of the interesting processes seem to be buried in
the background. The background is a general problem in $\gamma \gamma
\to $ hadrons physics.

The dispersive approach~\cite{PaukVanderhaeghen2013,Colangelo:2014pva}
is able to allow for real progress since contributions which we have
treated so far as separate contributions will be treated in an
integral manner. An example is the $\gamma \gamma \to
\pi\pi$ process which includes contributions attributed to the two--pion channel
, the pion--loop, the scalar contribution as well as the tensor
contribution. All-in-one can be gotten from the experimental data (see
e.g. figure~3 of~\cite{Jegerlehner:2013sja}). This also will settle such issues
as the pion polarizability. A lot remains to be done while new $a^{\rm
exp}_\mu$ is expected soon. For details see the Contribution by M.~Procura.\\

\acknowledgement
Thanks to the organizers for the kind invitation and support to the
``KLOE-2 Workshop on $e^+e^-$ collision physics at 1 GeV'' at
Frascati, and for giving me the opportunity to present this talk.


\begin{thebibliography}{}
%
%
%
%
%
%

\bibitem{Jegerlehner:2015stw}
  F.~Jegerlehner,
  EPJ Web Conf.\  {\bf 118}, 01016 (2016)

\bibitem{BNLfinal}
G.~W.~Bennett et al.  [Muon g-2 Collaboration],
Phys.\ Rev.\ D {\bf 73}, 072003 (2006)

\bibitem{Hertzog15}
  D.~W.~Hertzog,
  EPJ Web Conf.\  {\bf 118}, 01015 (2016)

\bibitem{HKS95}
M.~Hayakawa, T.~Kinoshita, A.~I.~Sanda,
Phys.\ Rev.\ Lett.\  {\bf 75}, 790 (1995);
Phys.\ Rev.\ D {\bf 54}, 3137 (1996);
M.~Hayakawa, T.~Kinoshita,
Phys.\ Rev.\ D {\bf 57}, 465 (1998)
[Erratum-ibid.\ D {\bf 66}, 019902 (2002)];

\bibitem{BPP1995}
J.~Bijnens, E.~Pallante, J.~Prades,
Phys.\ Rev.\ Lett.\  {\bf 75}, 1447 (1995)
[Erratum-ibid.\  {\bf 75}, 3781 (1995)];
Nucl.\ Phys.\ B {\bf 474}, 379 (1996);
[Erratum-ibid.\ {\bf 626}, 410 (2002)];\\
J.~Bijnens, J.~Prades,
Mod.\ Phys.\ Lett.\  A {\bf 22}, 767 (2007)

\bibitem{KnechtNyffeler01}
M.~Knecht, A.~Nyffeler,
Phys.\ Rev.\ D {\bf 65}, 073034 (2002)

\bibitem{Benayoun:2011mm}
M.~Benayoun, P.~David, L.~DelBuono, F.~Jegerlehner,
  Eur.\ Phys.\ J.\ C {\bf 72}, 1848 (2012);
%
 {\bf 73}, 2453 (2013)

\bibitem{Colangelo:2014pva}
  G.~Colangelo et al.,
  Phys.\ Lett.\ B {\bf 738}, 6 (2014)

\bibitem{KPPdeR02}
M.~Knecht, S.~Peris, M.~Perrottet, E.~de Rafael,
JHEP {\bf 0211}, 003 (2002)

\bibitem{MV03}
K.~Melnikov, A.~Vainshtein,
Phys.\ Rev.\ D {\bf 70}, 113006 (2004)

\bibitem{CMV03}
A.~Czarnecki, W.~J.~Marciano, A.~Vainshtein,
Phys.\ Rev.\ D {\bf 67}, 073006 (2003) [Erratum-ibid.\  D {\bf 73}, 119901 (2006)]

\bibitem{CMD203}
R.~R.~Akhmetshin et al.  [CMD-2 Collab.],
Phys.\ Lett.\ B {\bf 578}, 285 (2004);
%
V.~M.~Aulchenko et al.  [CMD-2 Collab.],
JETP Lett.\  {\bf 82}, 743 (2005)
[Pisma Zh.\ Eksp.\ Teor.\ Fiz.\  {\bf 82}, 841 (2005)];
R.~R.~Akhmetshin et al.,
JETP Lett.\  {\bf 84}, 413 (2006)
[Pisma Zh.\ Eksp.\ Teor.\ Fiz.\  {\bf 84}, 491 (2006)];
Phys.\ Lett.\  B {\bf 648}, 28 (2007)

\bibitem{SND06}
  M.~N.~Achasov et al. [SND Collab.],
  J.\ Exp.\ Theor.\ Phys.\  {\bf 103}, 380 (2006)
   [Zh.\ Eksp.\ Teor.\ Fiz.\  {\bf 130}, 437 (2006)]

\bibitem{KLOE08}
  A.~Aloisio et al. [KLOE Collab.],
  Phys.\ Lett.\  B {\bf 606}, 12 (2005);
  F.~Ambrosino et al. [KLOE Collab.],
  Phys.\ Lett.\  B {\bf 670}, 285 (2009)

\bibitem{KLOE10}
  F.~Ambrosino et al.  [KLOE Collab.],
  Phys.\ Lett.\ B {\bf 700}, 102 (2011)

\bibitem{KLOE12}
  D.~Babusci et al.  [KLOE Collab.],
  Phys.\ Lett.\ B {\bf 720}, 336 (2013)

\bibitem{BABARpipi}
  B.~Aubert et al.  [BABAR Collab.],
  Phys.\ Rev.\ Lett.\  {\bf 103}, 231801 (2009);
  J.~P.~Lees et al.,
  Phys.Rev. {\bf D86}, 032013 (2012)

\bibitem{BESIII}
  M.~Ablikim et al. [BESIII Collab.],
  Phys.\ Lett.\ B {\bf 753}, 629 (2016)

\bibitem{BaBar05}
B.~Aubert et al. [BABAR Collab.],
Phys.\ Rev.\ D {\bf 70}, 072004 (2004);
{\bf 71}, 052001 (2005);
{\bf 73}, 012005 (2006);
{\bf 73}, 052003 (2006);
{\bf 76}, 012008, ibid. 092005, ibid. 092006, (2007);
{\bf 77}, 092002 (2008)


\bibitem{BaBar11}
  J.~P.~Lees et al. [BaBar Collab.],
  Phys.\ Rev.\ D {\bf 85}, 112009 (2012);
%
{\bf 86}, 012008 (2012);
%
{\bf 87}, 092005 (2013);
%
{\bf 88}, 032013 (2013);
%
{\bf 89}, 092002 (2014)

\bibitem{Davier:2015bka}
M.~Davier,
Nucl.\ Part.\ Phys.\ Proc.\ {\bf 260}, 102 (2015)

\bibitem{Davier:2016udg}
  M.~Davier, A.~H\"ocker, B.~Malaescu, Z.~Zhang,
  Adv.\ Ser.\ Direct.\ High Energy Phys.\  {\bf 26}, 129 (2016)


\bibitem{Akhmetshin:2013xc}
  R.~R.~Akhmetshin et al. [CMD-3 Collaboration],
  Phys.\ Lett.\ B {\bf 723}, 82 (2013);
%
{\bf 759}, 634 (2016)

\bibitem{Kozyrev:2016raz}
  E.~A.~Kozyrev et al. [CMD-3 Collaboration],
  Phys.\ Lett.\ B {\bf 760}, 314 (2016)

\bibitem{Achasov:2014ncd}
  M.~N.~Achasov et al.,
  Phys.\ Rev.\ D {\bf 90}, no.11,  112007 (2014)

\bibitem{Aulchenko:2014vkn}
  V.~M.~Aulchenko et al. [SND Collaboration],
  Phys.\ Rev.\ D {\bf 91}, no.5,  052013 (2015)

\bibitem{Achasov:2016bfr}
  M.~N.~Achasov et al. [SND Collaboration],
  Phys.\ Rev.\ D {\bf 93}, no.9,  092001 (2016);
%
{\bf 94}, no.3,  032010 (2016);
%
{\bf 94}, no.9,  092002 (2016);
%
{\bf 94}, no.11,  112006 (2016);
%
{\bf 94}, no.11,  112001 (2016)

\bibitem{BES02}
J.~Z.~Bai et al.  [BES Collab.],
Phys.\ Rev.\ Lett.\  {\bf 84}, 594 (2000);
Phys.\ Rev.\ Lett.\  {\bf 88}, 101802 (2000);
  M.~Ablikim et al.,
  Phys.\ Lett.\ B {\bf 677}, 239 (2009)

\bibitem{Anashin:2015woa}
  V.~V.~Anashin et al.,
  Phys.\ Lett.\ B {\bf 753}, 533 (2016);
%
  arXiv:1610.02827 [hep-ex].


\bibitem{NLO}
B.~Krause,
Phys.\ Lett.\ B {\bf 390}, 392 (1997)

\bibitem{NNLO}
A.~Kurz, T.~Liu, P.~Marquard, M.~Steinhauser,
Phys.Lett. {\bf B734}, 144 (2014)

\bibitem{LbLNLO}
G.~Colangelo et al.,
Phys.Lett. {\bf B735}, 90 (2014)

\bibitem{Anastasi2016bvn}
  A.~Anastasi et al. [KLOE-2 Collaboration],
  doi:10.1016/j.physletb.2016.12.016
  arXiv:1609.06631 [hep-ex].

\bibitem{ALEPH}
R.~Barate et al. [ALEPH Collab.],
Z.\ Phys.\ C {\bf 76}, 15 (1997);
Eur.\ Phys.\ J.\ C {\bf 4}, 409 (1998);
S.~Schael et al. [ALEPH Collab.],
Phys.\ Rept.\  {\bf 421}, 191 (2005)

\bibitem{AlephCorr}
M.~Davier et al.,
Eur.Phys.J. {\bf C74}, 2803 (2014)

\bibitem{OPAL}
K.~Ackerstaff et al. [OPAL Collab.],
Eur.\ Phys.\ J.\ C {\bf 7}, 571 (1999)

\bibitem{CLEO}
S.~Anderson et al. [CLEO Collab.],
Phys.\ Rev.\ D {\bf 61}, 112002 (2000)

\bibitem{Belle}
  M.~Fujikawa et al.  [Belle Collab.],
  Phys.\ Rev.\ D {\bf 78}, 072006 (2008)

\bibitem{Benayoun:2015gxa}
  M.~Benayoun, P.~David, L.~DelBuono, F.~Jegerlehner,
  Eur.\ Phys.\ J.\ C {\bf 75}, no.12,  613 (2015)

\bibitem{JS11}
  F.~Jegerlehner, R.~Szafron,
  Eur.\ Phys.\ J.\ C {\bf 71}, 1632 (2011)

\bibitem{Davier:2010nc}
  M.~Davier, A.~H\"ocker, B.~Malaescu, Z.~Zhang,
  Eur.\ Phys.\ J.\ C {\bf 71}, 1515 (2011)
   [Erratum-ibid.\ C {\bf 72}, 1874 (2012)]

\bibitem{Davier:2009ag}
  M.~Davier et al.,
  Eur.\ Phys.\ J.\  C {\bf 66}, 127 (2010)

\bibitem{HMNT11}
  K.~Hagiwara, R.~Liao, A.~D.~Martin, D.~Nomura, T.~Teubner,
  J.\ Phys.\ G G {\bf 38}, 085003 (2011)

\bibitem{Davier:2016iru}
  M.~Davier,
  arXiv:1612.02743 [hep-ph].

\bibitem{Zhang:2015yfi}
  Z.~Zhang,
  EPJ Web Conf.\  {\bf 118}, 01036 (2016)
  doi:10.1051/epjconf/201611801036

\bibitem{JN}
  F.~Jegerlehner, A.~Nyffeler,
  Phys.\ Rept.\  {\bf 477}, 1 (2009)

\bibitem{PdRV}
  J.~Prades, E.~de Rafael, A.~Vainshtein,
  Adv.\ Ser.\ Direct.\ High Energy Phys.\  {\bf 20}, 303 (2009)

\bibitem{Feng:2011zk}
  X.~Feng, K.~Jansen, M.~Petschlies, D.~B.~Renner,
  Phys.\ Rev.\ Lett.\  {\bf 107}, 081802 (2011)

\bibitem{Golterman:2014ksa}
  M.~Golterman, K.~Maltman, S.~Peris,
  Phys.\ Rev.\ D {\bf 90}, no.7,  074508 (2014)
%

\bibitem{Borsanyi:2016lpl}
  S.~Borsanyi et al.,
  arXiv:1612.02364 [hep-lat].

\bibitem{Feng:2013xsa}
  X.~Feng et al.,
  Phys.\ Rev.\ D {\bf 88}, 034505 (2013)

\bibitem{Boyle:2011hu}
  P.~Boyle, L.~Del Debbio, E.~Kerrane, J.~Zanotti,
  Phys.\ Rev.\ D {\bf 85}, 074504 (2012)

\bibitem{Aubin:2013daa}
  C.~Aubin, T.~Blum, M.~Golterman, S.~Peris,
  Phys.\ Rev.\ D {\bf 88},  074505 (2013)

\bibitem{Francis:2014dta}
  A.~Francis et al.,
  arXiv:1411.3031 [hep-lat]

\bibitem{Malak:2015sla}
  R.~Malak et al. [Budapest-Marseille-Wuppertal Collab.],
  PoS LATTICE {\bf 2014}, 161 (2015)

\bibitem{Chakraborty:2016mwy}
  B.~Chakraborty, C.~T.~H.~Davies, P.~G.~de Oliviera, J.~Koponen, G.~P.~Lepage,
  arXiv:1601.03071 [hep-lat].

\bibitem{Burger:2015hdi}
  F.~Burger, X.~Feng, K.~Jansen, M.~Petschlies, G.~Pientka, D.~B.~Renner,
  EPJ Web Conf.\  {\bf 118}, 01029 (2016)

\bibitem{Burger:2013jya}
  F.~Burger et al. [ETM Collab.],
  JHEP {\bf 1402}, 099 (2014)

\bibitem{Aubin:2006xv}
  C.~Aubin, T.~Blum,
  Phys.\ Rev.\ D {\bf 75}, 114502 (2007)

\bibitem{DellaMorte:2016izp}
  M.~Della Morte, G.~Herdoiza, H.~Horch, B.~J\"ager, H.~Meyer, H.~Wittig,
  PoS LATTICE {\bf 2015}, 111 (2015)
  [arXiv:1602.03976 [hep-lat]]

\bibitem{DellaMorte:2011aa}
  M.~Della Morte, B.~J\"ager, A.~J\"uttner, H.~Wittig,
  JHEP {\bf 1203}, 055 (2012)

\bibitem{Calame:2015fva}
  C.~M.~Carloni Calame, M.~Passera, L.~Trentadue, G.~Venanzoni,
  Phys.\ Lett.\ B {\bf 746}, 325 (2015)

\bibitem{Abbiendi:2016xup}
  G.~Abbiendi et al.,
  arXiv:1609.08987 [hep-ex]

\bibitem{Jegerlehner:2008rs}
  F.~Jegerlehner,
  Nucl.\ Phys.\ Proc.\ Suppl.\  {\bf 181-182}, 135 (2008);
\textit{Hadronic effects in $(g - 2)_\mu$ and $\alpha_{\rm QED}(M_Z)$: Status and perspectives},
In: {\it Radiative Corrections},
 ed by J.~Sol\`a (World Scientific, Singapore 1999) pp 75--89
  [hep-ph/9901386].

\bibitem{Nyffeler:2016xul}
  A.~Nyffeler,
  EPJ Web Conf.\  {\bf 118}, 01024 (2016)

\bibitem{PaukVanderhaeghen2013}
  V.~Pauk, M.~Vanderhaeghen,
  Eur.\ Phys.\ J.\ C {\bf 74}, 3008 (2014)

\bibitem{FJ14}
F. Jegerlehner, Talk at the MITP Workshop
``Hadronic contributions to the muon anomalous magnetic moment'',
1-5 April 2014, Waldthausen Castle near Mainz,

\bibitem{Gerardin:2016cqj}
  A.~G\'erardin, H.~B.~Meyer, A.~Nyffeler,
  arXiv:1607.08174 [hep-lat]

\bibitem{Bijnens:2016hgx}
  J.~Bijnens, J.~Relefors,
  arXiv:1608.01454 [hep-ph].

\bibitem{Gnendiger:2013pva}
  C.~Gnendiger, D.~St\"ockinger, H.~St\"ockinger-Kim,
  Phys.\ Rev.\ D {\bf 88}, 053005 (2013)

\bibitem{Actis:2010gg}
  S.~Actis et al.,
  Eur.\ Phys.\ J.\ C {\bf 66}, 585 (2010)

\bibitem{CZYZ:2014yra}
  H.~Czy\.z,
  Int.\ J.\ Mod.\ Phys.\ Conf.\ Ser.\  {\bf 35}, 1460402 (2014)
  doi:10.1142/S2010194514604025

\bibitem{Jegerlehner:2017kke}
  F.~Jegerlehner, K.~Ko\l odziej,
  arXiv:1701.01837 [hep-ph].

\bibitem{Babusci:2011bg}
  D.~Babusci et al.,
  Eur.\ Phys.\ J.\ C {\bf 72}, 1917 (2012)

\bibitem{Nyffeler:2013oca}
  A.~Nyffeler,
  PoS CD {\bf 12}, 045 (2013)
  [arXiv:1306.5987 [hep-ph]]

\bibitem{Mennessier:1982fk}
  G.~Mennessier,
  Z.\ Phys.\ C {\bf 16}, 241 (1983)

\bibitem{Moussallam:2013una}
  B.~Moussallam,
  Eur.\ Phys.\ J.\ C {\bf 73}, 2539 (2013)

\bibitem{Pennington14}
  L.~Y.~Dai, M.~R.~Pennington,
  Phys.\ Rev.\ D {\bf 90}, no.3,  036004 (2014)

\bibitem{Jegerlehner:2013sja}
  F.~Jegerlehner,
  Acta Phys.\ Polon.\ B {\bf 44}, no.11,  2257 (2013)
\end{thebibliography}
\end{document}